% Revised version March 6 , 2007

\input harvmac.tex
\input epsf.tex

% \draftmode

\def\figin{\epsfcheck\figin}\def\figins{\epsfcheck\figins}
\def\epsfcheck{\ifx\epsfbox\UnDeFiNeD
\message{(NO epsf.tex, FIGURES WILL BE IGNORED)}
\gdef\figin##1{\vskip2in}\gdef\figins##1{\hskip.5in}% blank space instead
\else\message{(FIGURES WILL BE INCLUDED)}%
\gdef\figin##1{##1}\gdef\figins##1{##1}\fi}
\def\DefWarn#1{}
\def\figinsert{\goodbreak\midinsert}
\def\ifig#1#2#3{\DefWarn#1\xdef#1{fig.~\the\figno}
\writedef{#1\leftbracket fig.\noexpand~\the\figno}%
\figinsert\figin{\centerline{#3}}\medskip\centerline{\vbox{\baselineskip12pt
\advance\hsize by -1truein\noindent\footnotefont{\bf Fig.~\the\figno:} #2}}
\bigskip\endinsert\global\advance\figno by1}

%\KloseDD
\lref\frezar{
  T.~Klose and K.~Zarembo,
   ``Bethe ansatz in stringy sigma models,''
  J.\ Stat.\ Mech.\  {\bf 0605}, P006 (2006)
  [arXiv:hep-th/0603039].
  %%CITATION = HEP-TH 0603039;%%
}

\lref\bphase{
  N.~Beisert,
   ``On the scattering phase for AdS(5) x S**5 strings,''
  arXiv:hep-th/0606214.
  %%CITATION = HEP-TH 0606214;%%
 }
\lref\roibanklose{
 T.~Klose, T.~McLoughlin, R.~Roiban and K.~Zarembo,
   ``Worldsheet scattering in AdS(5) x S**5,''
  arXiv:hep-th/0611169.
  %%CITATION = HEP-TH 0611169;%%
  }

\lref\bes{
N.~Beisert, B.~Eden and M.~Staudacher,
   ``Transcendentality and crossing,''
  arXiv:hep-th/0610251.
  %%CITATION = HEP-TH 0610251;%%
}
\lref\bds{
  N.~Beisert, V.~Dippel and M.~Staudacher,
   ``A novel long range spin chain and planar N = 4 super Yang-Mills,''
  JHEP {\bf 0407}, 075 (2004)
  [arXiv:hep-th/0405001].
  %%CITATION = HEP-TH 0405001;%%
}

\lref\beisertnote{ N.~Beisert,
   ``Spin chain for quantum strings,''
  Fortsch.\ Phys.\  {\bf 53}, 852 (2005)
  [arXiv:hep-th/0409054].
  %%CITATION = HEP-TH 0409054;%%
}

\lref\dixon{
Z.~Bern, M.~Czakon, L.~J.~Dixon, D.~A.~Kosower and V.~A.~Smirnov,
   ``The four-loop planar amplitude and cusp anomalous dimension in maximally
  supersymmetric Yang-Mills theory,''
  arXiv:hep-th/0610248.
  %%CITATION = HEP-TH 0610248;%%
  }
\lref\bhl{ N.~Beisert, R.~Hernandez and E.~Lopez,
  ``A crossing-symmetric phase for AdS(5) x S**5 strings,''
  arXiv:hep-th/0609044.
  %%CITATION = HEP-TH 0609044;%%
}

\lref\sv{ M.~Spradlin and A.~Volovich,
  ``Dressing the giant magnon,''
  JHEP {\bf 0610}, 012 (2006)
  [arXiv:hep-th/0607009].
  %%CITATION = HEP-TH 0607009;%%
  C.~Kalousios, M.~Spradlin and A.~Volovich,
  ``Dressing the giant magnon. II,''
  arXiv:hep-th/0611033.
  %%CITATION = HEP-TH 0611033;%%
}
\lref\gm{
  P.~H.~Ginsparg and G.~W.~Moore,
  %``Lectures on 2-D gravity and 2-D string theory,''
  arXiv:hep-th/9304011.
  %%CITATION = HEP-TH 9304011;%%
}
\lref\hm{ D.~M.~Hofman and J.~M.~Maldacena,
  ``Giant magnons,''
  J.\ Phys.\ A {\bf 39}, 13095 (2006)
  [arXiv:hep-th/0604135].
  %%CITATION = HEP-TH 0604135;%%
}
\lref\fre{
 L.~D.~Faddeev and N.~Y.~Reshetikhin,
   ``Integrability Of The Principal Chiral Field Model In (1+1)-Dimension,''
  Annals Phys.\  {\bf 167}, 227 (1986).
  %%CITATION = APNYA,167,227;%%
}

\lref\janikun{ R. Janik, unpublished. }

\lref\coleman{S.~R.~Coleman and H.~J.~Thun,
   ``On The Prosaic Origin Of The Double Poles In The Sine-Gordon S Matrix,''
  Commun.\ Math.\ Phys.\  {\bf 61}, 31 (1978).
  %%CITATION = CMPHA,61,31;%%
}

 \lref\hl{
  R.~Hernandez and E.~Lopez,
   ``Quantum corrections to the string Bethe ansatz,''
  JHEP {\bf 0607}, 004 (2006)
  [arXiv:hep-th/0603204].
  %%CITATION = HEP-TH 0603204;%%
}

\lref\mt{ R.~R.~Metsaev and A.~A.~Tseytlin,
   ``Type IIB superstring action in AdS(5) x S(5) background,''
  Nucl.\ Phys.\ B {\bf 533}, 109 (1998)
  [arXiv:hep-th/9805028].
  %%CITATION = HEP-TH 9805028;%%
}
\lref\mannpolchtwo{
   N.~Mann and J.~Polchinski,
   ``Bethe ansatz for a quantum supercoset sigma model,''
  Phys.\ Rev.\ D {\bf 72}, 086002 (2005)
  [arXiv:hep-th/0508232].
  %%CITATION = HEP-TH 0508232;%%
}

\lref\mannpolchone{
  N.~Mann and J.~Polchinski,
   ``Finite density states in integrable conformal field theories,''
  arXiv:hep-th/0408162.
  %%CITATION = HEP-TH 0408162;%%
}

\lref\zamolodchikov{
  A.~B.~Zamolodchikov and A.~B.~Zamolodchikov,
   ``Factorized S-matrices in two dimensions as the exact solutions of  certain
   relativistic quantum field models,''
  Annals Phys.\  {\bf 120}, 253 (1979).
  %%CITATION = APNYA,120,253;%%
}

\lref\korepin{See for example,
V.E. Korepin, N.M. Bogoliubov, A.G. Izergin, ``Quantum inverse
 scattering method and correlation functions'',
Cambridge Univ. Press, 1993.
}

\lref\sosix{
  N.~Beisert, V.~A.~Kazakov and K.~Sakai,
   ``Algebraic curve for the SO(6) sector of AdS/CFT,''
  Commun.\ Math.\ Phys.\  {\bf 263}, 611 (2006)
  [arXiv:hep-th/0410253].
  %%CITATION = HEP-TH 0410253;%%
}

\lref\doreybound{ N.~Dorey,
   ``Magnon bound states and the AdS/CFT correspondence,''
  J.\ Phys.\ A {\bf 39}, 13119 (2006)
  [arXiv:hep-th/0604175].
  %%CITATION = HEP-TH 0604175;%%
 H.~Y.~Chen, N.~Dorey and K.~Okamura,
   ``Dyonic giant magnons,''
  JHEP {\bf 0609}, 024 (2006)
  [arXiv:hep-th/0605155].
  %%CITATION = HEP-TH 0605155;%% }
}

\lref\afs{ G.~Arutyunov, S.~Frolov and M.~Staudacher,
   ``Bethe ansatz for quantum strings,''
  JHEP {\bf 0410}, 016 (2004)
  [arXiv:hep-th/0406256].
  %%CITATION = HEP-TH 0406256;%%
}

\lref\gh{ C.~Gomez and R.~Hernandez,
   ``Integrability and non-perturbative effects in the AdS/CFT correspondence,''
  arXiv:hep-th/0611014.
}
\lref\beiserts{ N.~Beisert,
  ``The $su(2|2)$ dynamic S-matrix,''
  arXiv:hep-th/0511082.
  %%CITATION = HEP-TH 0511082;%%
}

\lref\magnonhopf{
  C.~Gomez and R.~Hernandez,
   ``The magnon kinematics of the AdS/CFT correspondence,''
  arXiv:hep-th/0608029.
  %%CITATION = HEP-TH 0608029;%%
  J.~Plefka, F.~Spill and A.~Torrielli,
   ``On the Hopf algebra structure of the AdS/CFT S-matrix,''
  Phys.\ Rev.\ D {\bf 74}, 066008 (2006)
  [arXiv:hep-th/0608038].
  %%CITATION = HEP-TH 0608038;%%
}

\lref\staudachers{ M.~Staudacher,
   ``The factorized S-matrix of CFT/AdS,''
  JHEP {\bf 0505}, 054 (2005)
  [arXiv:hep-th/0412188].
  %%CITATION = HEP-TH 0412188;%%
%\cite{Minahan:2002ve}
%\bibitem{Minahan:2002ve}
  J.~A.~Minahan and K.~Zarembo,
  ``The Bethe-ansatz for N = 4 super Yang-Mills,''
  JHEP {\bf 0303}, 013 (2003)
  [arXiv:hep-th/0212208].
  %%CITATION = HEP-TH 0212208;%%
}

\lref\krucz{
M.~Kruczenski, J.~Russo and A.~A.~Tseytlin,
   ``Spiky strings and giant magnons on S**5,''
  JHEP {\bf 0610}, 002 (2006)
  [arXiv:hep-th/0607044].
  %%CITATION = HEP-TH 0607044;%%
}

\lref\janikeq{
R.~A.~Janik,
   ``The AdS(5) x S**5 superstring worldsheet S-matrix and crossing symmetry,''
  Phys.\ Rev.\ D {\bf 73}, 086006 (2006)
  [arXiv:hep-th/0603038].
}

\lref\frolovlc{
 S.~Frolov, J.~Plefka and M.~Zamaklar,
  ``The AdS(5) x S**5 superstring in light-cone gauge and its Bethe
  equations,''
  J.\ Phys.\ A {\bf 39}, 13037 (2006)
  [arXiv:hep-th/0603008].
  %%CITATION = HEP-TH 0603008;%%
}

\lref\frolovalgebra{
  G.~Arutyunov, S.~Frolov, J.~Plefka and M.~Zamaklar,
   ``The off-shell symmetry algebra of the light-cone AdS(5) x S**5
   superstring,''
  arXiv:hep-th/0609157.
  %%CITATION = HEP-TH 0609157;%%
}

\lref\swansonetal{
  C.~G.~Callan, Jr., T.~McLoughlin and I.~J.~Swanson,
   ``Higher impurity AdS/CFT correspondence in the near-BMN limit,''
  Nucl.\ Phys.\ B {\bf 700}, 271 (2004)
  [arXiv:hep-th/0405153].
  %%CITATION = HEP-TH 0405153;%%
  C.~G.~Callan, Jr., T.~McLoughlin and I.~J.~Swanson,
   ``Holography beyond the Penrose limit,''
  Nucl.\ Phys.\ B {\bf 694}, 115 (2004)
  [arXiv:hep-th/0404007].
  %%CITATION = HEP-TH 0404007;%%
   C.~G.~Callan, Jr., H.~K.~Lee, T.~McLoughlin, J.~H.~Schwarz, I.~J.~Swanson and X.~Wu,
   ``Quantizing string theory in AdS(5) x S**5: Beyond the pp-wave,''
  Nucl.\ Phys.\ B {\bf 673}, 3 (2003)
  [arXiv:hep-th/0307032].
  %%CITATION = HEP-TH 0307032;%%
    T.~McLoughlin and I.~J.~Swanson,
  ``N-impurity superstring spectra near the pp-wave limit,''
  Nucl.\ Phys.\ B {\bf 702}, 86 (2004)
  [arXiv:hep-th/0407240].
  %%CITATION = HEP-TH 0407240;%%
  I.~J.~Swanson,
  ``Superstring holography and integrability in AdS(5) x S**5,''
  arXiv:hep-th/0505028.
  %%CITATION = HEP-TH 0505028;%%
}

\lref\bmn{
D.~Berenstein, J.~M.~Maldacena and H.~Nastase,
   ``Strings in flat space and pp waves from N = 4 super Yang Mills,''
  JHEP {\bf 0204}, 013 (2002)
  [arXiv:hep-th/0202021].
  %%CITATION = HEP-TH 0202021;%%
}

% \lref\hmpoles{D.~Hofman and J.~Maldacena, to appear. }

%%%%%%%%%%%%%%%%%%%%%%%%%%%%%%%%%%%%%%%%%%%%%%%%%%%%%%%%%%%%%%%%%

% \draftmode

\Title{\vbox{\baselineskip12pt  \hbox{ hep-th/0612079}}}
{\vbox{\centerline{Connecting giant magnons to the pp-wave: }
\centerline{ An interpolating limit of $AdS_5 \times S^5$ } }}
\bigskip
\centerline{ Juan Maldacena and Ian Swanson }
\bigskip
\centerline{ \it  School of Natural Sciences, Institute for Advanced Study}
\centerline{\it Princeton, NJ 08540, USA}

\vskip .3in
\noindent
%%%%%%%%%%%%%%%%%%%%%%%%%%%%%%%%%%%%%%%%%%%%%%%%%%%%%%%%%%%%%%%%%%%%%%%%%%%%%%%%%%%%%%%%%%%%
We consider a particular large-radius limit of the worldsheet $S$-matrix for
strings propagating on $AdS_5 \times S^5$.  This limiting theory interpolates
smoothly between the so-called plane-wave and giant-magnon regimes of the theory.
The sigma model in this region simplifies; it stands as a toy model of the full theory,
and may be easier to solve directly.  The $S$ matrix of the limiting theory is non-trivial,
and receives contributions to all orders in the $\alpha'$ expansion.
We analyze a guess for the full worldsheet $S$ matrix that was formulated recently by
Beisert, Hernandez and Lopez, and Beisert, Eden, and Staudacher,
and take the corresponding limit.  After doing a Borel resummation we find that the proposed
$S$ matrix reproduces the expected results in the giant-magnon region.
In addition, we rely on general considerations to draw some basic
conclusions about the analytic structure of the $S$ matrix.
%%%%%%%%%%%%%%%%%%%%%%%%%%%%%%%%%%%%%%%%%%%%%%%%%%%%%%%%%%%%%%%%%%%%%%%%%%%%%%%%%%%%%%%%%%%%

 \Date{ }

%%%%%%%%%%%%%%%%%%%%%%%%%%%%%%%%%%%%%%%%%%%%%%%%%%%%%%%%%%%%%%%%%%%%%%%%%
\newsec{Introduction}
\noindent
%%%%%%%%%%%%%%%%%%%%%%%%%%%%%%%%%%%%%%%%%%%%%%%%%%%%%%%%%%%%%%%%%%%%%%%%%
Recently there has been a lot of activity regarding the worldsheet $S$ matrix for type IIB
string theory on $AdS_5 \times S^5$,
and the corresponding $S$ matrix\foot{This $S$ matrix should not be confused with the
notion of a {\it spacetime} $S$ matrix.} for planar ${\cal N}=4$ super-Yang Mills
theory.  This is an object that arises when one considers
 operators or states with very large charge $J$ under an $SO(2)$
subgroup of $SO(6)$.  In the limit $J\to \infty$ with $\Delta - J$ finite, where
$\Delta$ is the conformal dimension, we are led to consider a finite set of
impurities propagating along an infinite spin chain \staudachers.
The Hamiltonian of this spin chain is formulated to reproduce the action of the
gauge theory dilatation generator on single-trace operators in ${\cal N}=4$ super-Yang
Mills theory.
On the string theory side, one can go to light-cone gauge and obtain
a rather complicated looking two dimensional theory on an infinite line
\refs{\mt,\frolovlc}.
Both systems have elementary excitations that have been dubbed magnons.
These theories are conjectured to be integrable, so that
the full $S$ matrix on either side of the correspondence
is determined entirely in terms of $2\to 2$
scattering processes.  It was shown by Beisert that the {\it matrix } structure of the
basic $2 \to 2 $ $S$ matrix is fixed by symmetries \beiserts, so what remains is to determine
the phase factor.  Recently there has been an interesting guess for this phase proposed by
Beisert, Eden and Staudacher \bes, based on previous work by Beisert, Hernandez and Lopez \bhl.
We will refer to this guess as BES/BHL.\foot{The paper \bhl\ contains a
couple of guesses. The correct one, based on the results in \dixon, seems to be the one in \bes.}

In this article we study an interesting strong-coupling limit of the worldsheet $S$ matrix,
 wherein
the sigma model describing the system simplifies, but the $S$ matrix itself remains non-trivial.
We find this limit interesting in that it still captures some important aspects of the full
 problem.
As we take $\lambda \equiv g_{\rm YM}^2 N  \to \infty$, we can scale the magnon momentum
$p$ ($p\sim p + 2 \pi$) in different ways \bhl.
 The simplest limit is taken by scaling $p$ such that $p \sqrt{\lambda } =$ fixed. This
produces the plane-wave limit of \bmn, where excitations are free and the $S$ matrix is unity.
Another simple limit is to keep $p$ fixed. In this case the elementary excitations can be
viewed as non-topological solitons of a weakly coupled two-dimensional theory.
Here, the basic magnon excitations of the theory turn into large solitons, or ``giant magnons'' \hm.
The giant-magnon $S$ matrix can be computed
using semiclassical methods, and one obtains an answer that
scales as $\log S \sim \sqrt{\lambda} f(p_1,p_2)$, where the function $f$ has a branch cut
at $p_1 =p_2$. In the exact theory we expect that this branch cut is replaced by a sequence
 of poles or zeros with a spacing of order $1/\sqrt{\lambda}$.
In both of these limits the leading-order answer for the $S$ matrix is described by a weakly
coupled theory that can be solved easily.

The third interesting limit corresponds to scaling $p$ in such a way that one probes the region
in between
the previous two regimes.  Namely, we keep $p \lambda^{1/4}$ fixed. In this
regime the $S$ matrix is non-trivial and receives contributions to all orders in $\alpha'$.
The $S$ matrix develops a singularity for complex momenta that approaches the real axis for
large $\lambda$.  After going to suitable rescaled variables, however, the singularity
remains a finite distance from the axis.  This singularity is intimately related to the
structure Janik's crossing-symmetry equation \refs{\beiserts,\janikeq}, which remains
 non-trivial even after taking this limit.
 Furthermore, the full sigma model describing strings in $AdS_5\times S_5$ simplifies significantly
in this limit, and leads (after gauge fixing)
 to a rather simple-looking (albeit non Lorentz-invariant) theory in $1+1$ dimensions.
The magnons in this theory have rescaled momenta that lie between zero and infinity.
For momenta close to zero, the $S$ matrix becomes the identity and we recover the plane-wave
results as well as the leading finite-$J$ deviation away from the plane-wave limit \swansonetal.
For large rescaled
momenta, however, the elementary excitations turn into solitons, and we recover results
that are similar to those obtained for the giant-magnon regime in \hm. In particular,
we find a semiclassical $S$ matrix with a branch cut, which should be replaced by a string of
poles in the exact answer.

The limit characterized by $p \sim \lambda^{-1/4}$ was studied in \afs, 
as it arises when one considers the flat-space limit.
Our discussion is not directly relevant to the computation of energies of
string states in flat space, because the part of the $S$ matrix we consider drops out from that
computation.  We are interested in the particular part of the $S$ matrix we consider here
because it stands as a toy version of the full
$S$ matrix, and displays several interesting features of the full problem.  Moreover,
it is described by a self-contained Lagrangian that might prove easier to solve directly than
the full $AdS_5 \times S^5$ theory (though we were not able to solve it).

We can also consider the corresponding limit of the recent BES/BHL \refs{\bes,\bhl}
proposal for the $S$ matrix. After a Borel resummation,
the phase is given by a  simple-looking integral expression that allows us to
explore some of its analytic properties.  In particular, we show that we recover the expected
structure in the giant-magnon region.  Namely, we see that the branch cut in the
semiclassical answer disappears and is replaced by a sequence of double poles.
We also check that the crossing-symmetry equation is obeyed
after choosing an appropriate contour.

The paper is organized as follows.  In Section 2 we discuss the kinematics of this
``near-flat-space'' limit.
In Section 3 we discuss the worldsheet theory that is obtained in this limit.  We first
consider an analogous limit for the $O(N)$ sigma model, and then we turn to the full
$AdS_5 \times S_5$ theory.  For each case we find a pair of theories
describing the system before or after imposing the Virasoro constraints.
In Section 3 we consider the BES/BHL  $S$ matrix in this limit and study some of its
properties.  We also include several appendices where we discuss various related topics.

%%%%%%%%%%%%%%%%%%%%%%%%%%%%%%%%%%%%%%%%%%%%%%%%%%%%%%%%%%%%%%%%%%%%%%%%%%%%%%%%%%%%%%%%%%%
\newsec{The ``near-flat-space'' limit}
\noindent
%%%%%%%%%%%%%%%%%%%%%%%%%%%%%%%%%%%%%%%%%%%%%%%%%%%%%%%%%%%%%%%%%%%%%%%%%%%%%%%%%%%%%%%%%%%
As mentioned in the introduction, we are interested in a limit in which we scale the
magnon momentum $p$ such that $p \sim \lambda^{-1/4}$.
This particular scaling choice can be motivated by introducing the kinematic variables used in \bds,
\beisertnote:
\eqn\kingv{
x^+ + { 1 \over x^+ } - x^- - { 1 \over x^-} = { i \over g}  ~,~~~~~~~~~~
e^{ip} = { x^+ \over x^-}~,~~~~~~~g^2 \equiv { \lambda \over 16 \pi^2} =
{ g_{\rm YM}^2 N \over 16 \pi^2 }\ ,
}
where $g_{\rm YM}$ is the conventional Yang Mills coupling.
Note that we  have replaced $p$ by a pair of variables $x^\pm$ obeying a constraint equation.
The matrix structure of the $S$ matrix
is dramatically simplified when expressed in terms of these variables \beiserts.
Furthermore, crossing symmetry acts in a rather simple fashion: $x^\pm \to 1/x^\pm $.

\ifig\dispersion{  The dispersion relation $\epsilon(p)$ at large $\lambda$. The plane-wave region
corresponds to the region near $p\sim 0$, where the dispersion relation looks like that of a
relativistic massive particle. In the giant-magnon region, the dispersion relation is essentially
$\epsilon \sim \sqrt{\lambda } \sin p/2$.  We are interested in the interpolating region, which overlaps
with limits of both of the previous regimes. }
{\epsfxsize4.0in\epsfbox{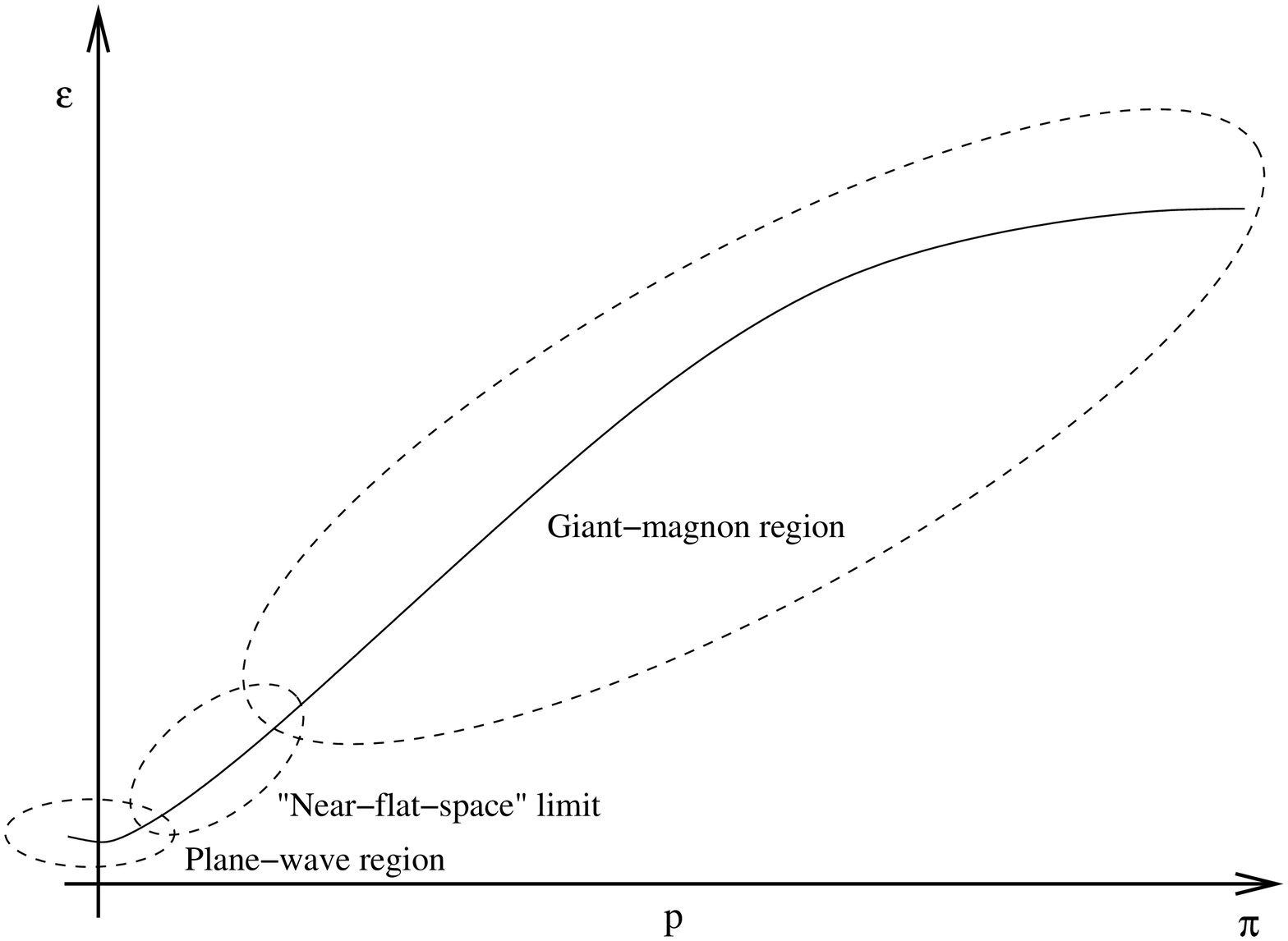}}

We now want to focus on the regime where $x^+ \sim x^- \sim 1$, as it connects the
two possible ways of approximately solving the equation in \kingv\ at strong coupling, namely,
$x^+ \sim  (x^-)^{\pm 1 }$.   We define rescaled variables $w^\pm$ via
\eqn\xpmdefw{
x^\pm = e^{ w^\pm/\sqrt{g}}  ~,
}
and take $g\to \infty$, keeping $w^\pm$ fixed.
In this limit we can see that the  constraint \kingv\ linking $x^+$ and $x^-$ becomes
\eqn\constrbe{
(w^+)^2 - (w^-)^2 = i ~,~~~~~~~~~~ k \equiv p \sqrt{g} = -i ( w^+ - w^-)\ ,
}
where we have defined a rescaled momentum $k$ that is kept fixed as we take the limit.
These equations can be solved to express $w^\pm$ in terms of $k$:
\eqn\wpmk{
w^\pm = { 1 \over 2 }
\left( { 1 \over k} \pm i  k \right)\ .
}
It will also be  useful to introduce a new variable $u$, defined as
\eqn\varu{
u \equiv { 1 \over 2} ((w^+)^2 + (w^-)^2 ) = (w^\pm)^2 \mp {i \over 2} = { 1 \over 4 } (
{ 1 \over k^2 } - k^2 )\ ,
}
where the momentum $k$ is positive.
Small $k$ ($k\ll 1$) corresponds  to the large-momentum regime of the plane-wave limit, while
large $k$ ($k\gg 1$) probes the small-momentum regime of the giant-magnon region:
see \dispersion.
In these two regimes we have a simple, weakly-coupled description of the dynamics.
For $k\sim 1$ we are forced to consider an interacting theory.

It will also be useful to consider the following expression for the energy:
\eqn\energin{
\epsilon = { g \over i } \left( x^+ -{ 1 \over x^+} -x^- + { 1 \over x^-} \right) = \sqrt{ 1 +
16 g^2 \sin^2{ p\over 2 }}\ .
}
To leading order in the strong-coupling expansion ($g \gg 0$), we find that the energy
is $\epsilon \sim 2 \sqrt{g} k $, representing
particles that move close to the speed of light to the right.
The two approximate solutions under consideration
correspond to the plane-wave and giant-magnon regimes.
Of course, there is a similar region where $x^+ \sim x^- \sim -1$, where we
get particles that move very fast to the left. This region is related by worldsheet
parity to the one discussed here.

It will be useful to  define a rescaled energy by picking
out the subleading $k$ dependence as
\eqn\rescen{
\hat \epsilon = \lim_{g \to \infty} \left[{2 \sqrt{g}} ( \epsilon - 2 {\sqrt{g}} k ) \right]=
{ 1 \over 2 k} - { k^3 \over 6  }\ .
}
Now that the dispersion relation is not exactly relativistic, we can see that excitations will
travel with different velocity, and one can define an $S$ matrix that encodes scattering
between two of these right-moving
excitations.  In other words, the velocity is
\eqn\velocit{
v = { d \epsilon \over d k } = 2 \sqrt{g} - { 1 \over 4 \sqrt{g} }
 ( { 1 \over k^2 } + k^2 )\ .
}
Physically, the rescaling of the energy in Eqn.~\rescen\ means that
it will take a long time for the magnons to separate, and
small deviations of the metric from flat space lead to large effects, which, in turn,
lead to a non-trivial $S$ matrix.

Let us pause here to consider
the relation between this limit and the flat-space limit considered in
\afs . When one is interested in reproducing the energies of strings in flat space, one should
rescale $J \sim \lambda^{1/4}$.  Strictly speaking, this is not the infinite-$J$ limit, and
one cannot be sure that the asymptotic $S$ matrix formulas apply.  Nevertheless, one can
proceed and find that the order of magnitude of the momentum,
$p \sim n/J \sim n/\lambda^{1/4}$, indeed scales as discussed above.
In this case, one can keep the leading dependence of the energy
$\epsilon \sim 2 \sqrt{g} k$, but consider both left- and right-movers to have zero total
momentum on the worldsheet.  Upon writing the Bethe equations, one needs only
the $S$ matrix between left- and
right-movers, but the $S$ matrix for the left-movers drops out of the analysis.
The $S$ matrix between left- and
right-movers is rather simple, and one can obtain the flat space results \afs\ in a
straightforward manner.
In summary, the discussion presented here will not be relevant for finding
the {\it energies} of states in flat space (which
was already done in \afs ). Rather, we will be primarily concerned with
understanding the full structure of the $S$ matrix and the connection
between the two simple strong-coupling regimes discussed above
(i.e.,~those of the plane wave and giant magnon).

We can therefore view our limit as a toy version of the full
problem, where we have retained some of the interesting structure
of the complete theory, but we have lost one parameter, namely
$\lambda$. In fact, one can check that in this limit the structure
of the $S$ matrix \beiserts\ remains nontrivial and, consequently,
the crossing symmetry equation \janikeq\ is also nontrivial and
becomes:\foot{The conventions we employ in defining $S_0$ and
$\sigma$ are the same as those in \bhl . } \eqn\crosseq{\eqalign{
S_0(-w_1,w_2) S_0(w_1,w_2) =\sigma^2(-w_1,w_2)& \sigma^2(w_1,w_2)
=  \left( {( w_1^- + w_2^-)(w_1^- - w_2^+ ) \over  (w_1^+ +
w_2^-)(w_1^+ - w_2^+)    }\right)^2\ , \cr {\rm where
~~~~}S_0(w_1,w_2) =& { (w_1^- - w_2^+)(w_1^+ + w_2^-)
    \over (w_1^+ - w_2^-)(w_1^- + w_2^+) }  \sigma^2\ .
}}
The crossing transformation itself maps $w^\pm \to - w^\pm$ along a path that we will specify later.
It is amusing to note that the constraint \constrbe\ looks very similar to the constraint between energy
and momentum in a relativistic theory in $1 + 1$ dimensions.
Thus one may introduce a rapidity variable $\eta$ via\foot{The system does not
have relativistic invariance under $\eta \to \eta \, + $ constant. For
example,
 the right-hand side of \crosseq\ is not simply a function of $\eta_1 - \eta_2$.}
\eqn\wpmfrometa{
w^+ = e^{ i \pi/4} \cosh \eta ~,~~~~~~~~w^- = e^{ i \pi /4} \sinh \eta\ .
}
In fact, the variable $\eta$, which starts out
living on a cylinder,
is what remains after taking
the near-flat-space limit of Janik's torus \janikeq .

One can attempt to solve the crossing equation directly by looking for a
meromorphic (but not periodic) function in $\eta$.
The right-hand side of the crossing
equation \crosseq , however, has
a structure that excludes any such solution.
(For a detailed argument see Appendix A.)
One is therefore forced to introduce branch cuts somewhere. In fact, the recently proposed solutions
in \bes, \bhl\ and \bphase\ all
have branch cuts, as we will see later.  A similar result is also true for the full crossing
symmetry equation presented in \janikeq . Namely, {\it there is no meromorphic solution as a function of
the coordinates of the torus $z_1, z_2$,} even after going to the covering spaces
of the two tori (which comprise two complex planes).

%%%%%%%%%%%%%%%%%%%%%%%%%%%%%%%%%%%%%%%%%%%%%%%%%%%%%%%%%%%%%%%%%%%%%%%%%%%%%%%%%%%%%%%%%
\newsec{Lagrangian in the near-flat-space limit}
\noindent
%%%%%%%%%%%%%%%%%%%%%%%%%%%%%%%%%%%%%%%%%%%%%%%%%%%%%%%%%%%%%%%%%%%%%%%%%%%%%%%%%%%%%%%%%
In this section we consider the near-flat-space limit of the sigma model.
The limiting sigma model is a well defined system on its own, which looks simpler than the
original system.  We investigate its properties with the hope that it will prove easier
than the full sigma model to solve directly.

%%%%%%%%%%%%%%%%%%%%%%%%%%%%%%%%%%%%%%
\subsec{The $O(N)$ case }
\noindent
%%%%%%%%%%%%%%%%%%%%%%%%%%%%%%%%%%%%%%
As an exercise, let us consider the $O(N)$ sigma model.  The target space of this sigma model
is a sphere $S^{N-1}$.  Let us consider a state with a constant spin density $J = J_{12}$, where
$J_{kl}$ are rotation generators in the $kl$ plane. We denote the corresponding angle on
the sphere by $\varphi$.
Starting with a classical analysis, we see that
the lowest energy state is given simply by $\dot \varphi =$ constant.
For simplicity, we can take the case with $N=3$ and parametrize the $S^2$
according to
\eqn\stwosig{
S = { R^2 \over 4 \pi } \int \cos^2 \theta ( \partial \varphi )^2  + (\partial \theta )^2
~,~~~~~~~~~{R^2 \over 4 \pi} = g\ .
}
Starting with a solution for which $\dot \varphi =1 $, $\theta =0$, we can
 perform a boost on the worldsheet coordinates
 $  \tilde \sigma^\pm = \tilde \sigma^0 \pm \tilde
 \sigma^1 $ and expand in small fluctuations around the
 constant-spin-density solution
\eqn\varresc{ \eqalign{
 \tilde  \sigma^+ &= 2 \sqrt{g} \ \sigma^+ , ~~~~~~~
 \tilde \sigma^- =  { \sigma^- \over 2 \sqrt{g} }\ ,
\cr
\varphi &= { \tilde \sigma^+ + \tilde \sigma^- \over 2 } + { \delta \over \sqrt{g} } =
  \sqrt{g}  \sigma^+ + { \chi \over \sqrt{g} }\ ,
  \cr
  \chi &= {  \sigma^- \over 4 } + \delta\ ,
\cr
 \theta &= { y \over  \sqrt{g}}\ ,
\cr
g &\to\infty\ ,
}}
where $\sigma^{\pm}$ are the coordinates after performing the boost.

Note that we will be interested in
solutions where $\chi = { 1 \over 4} \sigma^- + \delta$,  with $\delta$ representing
a small fluctuation.  Under these rescalings, the action, up to constant and total derivative terms,
is finite:
\eqn\confth{ S= 4
\int  \partial_- \chi \partial_+ \chi  + \partial_+ y \partial_-y -
 y^2 \partial_- \chi\ .
 }
By setting $\chi = { 1 \over 4} \sigma^- + \delta$, we see that to
leading order we have a massless field $\delta$ and a  massive field $y$.
Long wavelength fluctuations
in $\delta$ are simply sound waves, or spin-density waves  propagating along the system.
The massive field $y$ is the massive
field that arises in the plane-wave limit.
We thus find that the system is described by a
simple Lagrangian \confth\ in the near-flat-space limit.
  Notice that the only nontrivial interaction is
a cubic interaction that breaks Lorentz invariance.
 To generalize to the case of an $O(N)$ (as opposed to $O(3)$)
 sigma model, all we need to do is replace
$y^2 \to \vec y^{\,\,2}$ in the above Lagrangian, where $\vec y$ has $N-2$ components.

From the point of view of the $O(N)$ sigma model, this limit has the following interpretation.
The ground state with constant spin density gives rise to a sort of Fermi sea of particles, all
with the same spin. Since the particles are interacting, it is not exactly a Fermi sea, but in the
thermodynamic limit there is a sharp cutoff in the momentum of the particles \korepin .
The theory  described by \confth\ corresponds to considering excitations around the ground state
that are moving with a momentum close to one of the Fermi 
momenta\foot{From this point of view, this limit is reminiscent of the double scaling limit
of matrix models, where one focuses near the endpoint of a distribution of eigenvalues
(see, e.g.,~ref.~\gm\ for a review). } $\pm p_f$.

The equations of motion for the theory in Eqn.~\confth\ appear as
\eqn\eqmot{\eqalign{
0&=  \partial_- j_+ ~,~~~~~j_+ \equiv  \partial_+\chi -  { y^2 \over 2 }\ ,
\cr
0&=\partial_- \partial_+ y + y \partial_-\chi\ ,
}}
where the first equation implies the existence of a chiral conserved current $j_+$.
It is useful to understand what happens to the Virasoro generators
in this limit.  The $T_{--}$ generator is
\eqn\stresste{
T_{--} = ( \partial_- \chi )^2 + ( \partial_- y)^2\ ,
}
and it is conserved according to \eqmot :  $\partial_+ T_{--} =0$.
In fact, we see that the action \confth\ is right-conformal invariant.
In other words, it is invariant under $\sigma^- \to f(\sigma^-)$.  However,
the action is not invariant under left-moving conformal transformations.
The original left-moving conformal symmetry has become a
chiral $U(1)$ symmetry acting as
$\chi \to \chi +
 \epsilon(\sigma^+)$.
The right-moving stress tensor of the original theory takes the form
\eqn\strright{
T_{++} = { g \over 2 } + j_+ + o(1/g)\ ,
}
where $j_+$ was defined in \eqmot .

In the full theory we will impose Virasoro constraints after adding a timelike direction,
and we consider a solution with $\dot t =1$, or $t= \tilde \sigma^0 = \sqrt{g}
\sigma^+ + { \sigma^- \over 4 } $ (see \varresc ).
The leading-order term in \strright\  should
be equated with the contribution of the timelike direction
to the stress tensor.  The zeroth-order term in \strright\
should be set to zero, which leads to the constraint
$j_+=0$. Similarly, \stresste\
should be equated with the time-like contribution, which leads to
the constraint $T_{--} = {1 \over 16}$. We see that both constraints are attainable
classically, since they are related to symmetries of the theory.

After adding a trivial time direction and imposing the two Virasoro constraints
\eqn\virlim{
j_+=0 ~,~~~~~~~T_{--} = { 1 \over 16 }\ ,
}
we can move to a gauge-fixed Lagrangian by defining new coordinates
\eqn\newcor{
 x^+ \equiv \sigma^+ ~,~~~~~~~~x^- \equiv {  \sigma^- \over 2 }  +
  2  \chi = { \sigma^-   } + 2 \delta\ .
 }
 The derivatives then transform as
 \eqn\derivtr{\eqalign{
 \partial_{\sigma^+}  &=
  \partial_{x^+} + 2 ( \partial_{\sigma^+} \chi  ) \partial_{x^-} =
  \partial_{x^+} + { y^2   }  \partial_{x^-}\ ,
\cr
\partial_{\sigma^-} &= 2 ( { 1 \over 4 }  + \partial_-\chi ) \partial_{x^-}\ ,
}}
where we used the constraint $j_+=0$.
After using the Virasoro condition $T_{--}={ 1 \over 16}$,  one can check that the
 gauge-fixed Lagrangian becomes\foot{Essentially, we perform the change of variables in the
equations of motion and write the Lagrangian from which they follow.}
 \eqn\intlagr{
 S = 4
 \int dx^+ dx^-  \left[  \partial_+ y   \partial_- y - { 1 \over 4} y^2 +
  y^2  ( \partial_{-} y)^2 \right]\ .
 }
The momentum is given by
\eqn\totmom{\eqalign{
 k =& \sqrt{g} p = \sqrt{g} \int ( d \varphi - d \sigma^0) = \int d\chi - { 1 \over 4 } d\sigma^- =
 \int d\delta  \cr
  =&\int [ - 2 (\partial_- y )^2 dx^- + ( { y^2 \over 2 } + 2 y^2 (\partial_- y)^2 ) dx^+ ]\ ,
 } }
 where the derivatives in the second line are taken with respect to $x^\pm$.
Note that we can understand
 the momentum both as an angle\foot{Note that in this limit $\delta$ ceases to be periodic.} and 
as a Noether charge under $x^-$ translations.
In the limit of interest, the momentum $k$ is not periodic but is defined on the semi-infinite
 line $(0,\infty)$.
The expression for the energy takes the form
\eqn\energy{\eqalign{
-k_+ = & \, \hat \epsilon = -4  \int  [(\partial_+ y)^2  + ( \partial_+ \chi )^2 ] d\sigma^+   - y^2 d\chi
  \cr
 = & \, 4 \int  [  { y^2 \over 4 } - y^2 (\partial_- y)^2 ] dx^-
 -[ (\partial_+ y)^2 + 2 y^2 \partial_+y \partial_- y ] dx^+\ ,
 }}
where $\partial_+ = \partial_{\sigma^+}$ in the first line, and
$\partial_\pm = \partial_{x^\pm }$ in the second.

It is interesting to consider giant-magnon solutions to this Lagrangian. For this purpose
we consider the $O(N)$ case with $N\geq 4$, and we focus on the first two components
of $\vec y$.  We can start with the solution carrying an additional angular
momentum $J_2$ in the complex $y_1 + i y_2$
  plane \refs{\doreybound,\sv,\krucz}, and then take the limit.  We obtain
\eqn\rescaledgian{\eqalign{
\chi &= { 1 \over 4}  \sigma^- + \delta ~,~~~~~~~~\delta =  { k \over 2} \tanh u\ ,
\cr
y^1 + i y^2 &= e^{i v } { k \over 2  \cosh u }\ ,
\cr
 u &  \equiv   - { k^3 \over { J_2^2 } + k^4 } \sigma^- + {
k \over 4 }\sigma^+\ ,
\cr
v & \equiv   { J_2 k \over
{J_2^2  } + k^4  } \sigma^- + { J_2 \over 4 k }\sigma^+\ .
}}

To think about the {\it elementary} giant magnon we can set $J_2=1$;
the other solutions can be thought of as bound states of the elementary solution \doreybound .
The energy of these states is given by
\eqn\energs{
\hat \epsilon =  { J_2^2 \over 2 k } - {k^3 \over 6}\ ,
}
which, for $J_2=1$, reduces to \rescen .
Notice that the velocity of these particles can be
computed as
\eqn\veloc{\eqalign{
v_- =&  { d \sigma^- \over d\sigma^+} =- { d \hat \epsilon \over d k} =
    { 1 \over 2} ( { 1 \over k^2 } + k^2 )
\cr
 = & {d \sigma^1 \over d \sigma^0}
= { 1 - v_- \over 1 + v_-} = - { ( { 1 \over k } - k )^2 \over ({ 1 \over k } + k )^2 }\ .
}}
Notice also that they always move with a speed that is less than the speed of light, and
they all move to the left.\foot{Before performing the boost in Eqn.~\varresc ,
they traveled to the right \velocit .}

Near $k\sim 0$, the above theories \confth ,\ \intlagr\
are weakly coupled, while as $k$ increases toward the region
$k \sim 1$, the theories become strongly coupled.
We expect that the elementary
excitations become giant-magnon solutions \rescaledgian\ for large $k$.
In this latter region the giant magnon is extended, so that the classical
description becomes appropriate. In fact, we can see that the size of the solution
is of order $k$ for large $k$.
We can find the giant-magnon solution for the Lagrangian
\intlagr\ by defining the coordinates $x^-$ as above \newcor .
The equation is invertible since ${ d x^- \over d \sigma^-} = 1 + 2 \partial_-\delta>0$.

Because they are limits of integrable theories, the above Lagrangians  \confth ,\ \intlagr\
are themselves integrable. In Appendix B we display explicitly the Lax connection for
$O(N)$ theories in this limit.
We can ask whether the theory \confth\ remains right-conformal after we take into account quantum
corrections. In Appendix B we argue that right conformal symmetry is broken
if $N\not =2$, but for $N=2$ the theory
remains conformal.  Thus, the theories \confth\ and \intlagr\
will not be equivalent as quantum theories. We know, however, that the theory
\confth\ is quantum integrable, since it is a limit of the integrable $O(N)$ sigma model.
We could therefore solve the quantum theory \confth\
by taking a limit of the $O(N)$ quantum theory \zamolodchikov , but we will not
do so here. However, let us mention one result. If one computes the scattering amplitude
for giant magnons in the region where $ k_1 - k_2 \ll k_{1,2} \gg 1$, one finds that
the branch cut in the semiclassical scattering amplitude,
 which is the same for all $N$, becomes
a sequence of single poles for $N\not =2$ and a sequence of double poles for $N=2$. For
$N=2$, the model can be viewed as a limit of an $OSP(M+2|M)$ theory, as explained in
\mannpolchtwo .

The theory in Eqn.~\intlagr\ may or may not be integrable at the quantum level.
However, for $N=4$, \intlagr\ is also a limit of the Fadeev-Reshetikin theory \refs{\fre,\frezar},
so it must be quantum integrable in this case.
These theories are reminiscent of the chiral Potts
model, in that they are integrable and they break Lorentz symmetry.\foot{We thank 
B.~McCoy for a discussion on this topic.}

%%%%%%%%%%%%%%%%%%%%%%%%%%%%%%%%%%%%%%%%%%%%%%%%%%%%%%%%%%%%%%%%%%%%%%%%%%%%%%%%
\subsec{The near-flat-space limit of the $AdS_5 \times S^5$ sigma model}
\noindent
%%%%%%%%%%%%%%%%%%%%%%%%%%%%%%%%%%%%%%%%%%%%%%%%%%%%%%%%%%%%%%%%%%%%%%%%%%%%%%%%
We can now consider the full $AdS_5 \times S^5$ sigma model,
starting with the Green-Schwarz action as written in \mt .
We parametrize by $\varphi$ the angle on $S^5$ that is shifted by the action of the
generator $J$.  We also pick $t$ to be the coordinate on $AdS_5$ whose shift
corresponds on the field theory side to the action of the gauge theory
dilatation operator $\Delta $.  We then perform the following rescalings
\eqn\rescalings{\eqalign{
t =&  \sqrt{g} \sigma^+ + { \tau \over \sqrt{g} } ~,~~~~~~~\varphi = \sqrt{g} \sigma^+ + { \chi \over \sqrt{g}}
~,~~~~~~~~ \vec \theta = \vec y/\sqrt{g} ~,~~~~~~~\vec \rho = \vec z/\sqrt{g}\ ,
\cr
& \Theta^1 \sim  { \psi_- \over g^{1/4} } ~,~~~~~~~~~
\Theta^2 \sim  { \psi_+ \over g^{3/4} } ~, ~~~~~ g \to \infty\ ,
}}
where $y$ denotes the four transverse directions in the $S^5$ subspace, $z$ denotes
the transverse coordinates on the $AdS_5$ subspace, and $\Theta^i$ denote two ten-dimensional
Weyl spinors of type IIB string theory after fixing kappa symmetry (so that $\psi_\pm$ are
$SO(8)$ spinors).

Upon taking the limit we obtain the Lagrangian
\eqn\finlagrag{\eqalign{
{\cal L} = & 4 \left\{
- \partial_+ \tau \partial_- \tau +  \partial_+ \chi \partial_- \chi +
  \partial_+ \vec z \partial_- \vec z +  \partial_+ \vec y \partial_- \vec y  -
   {\vec y}^2  \partial_- \chi  - {\vec z}^2  \partial_-\tau \right.
\cr   &
+ i \psi_+  \partial_- \psi_+ + 2 i( \partial_- \tau + \partial_- \chi)
 \psi_- \partial_+ \psi_- + 2i ( \partial_- \tau + \partial_- \chi)  \psi_-
\Pi \psi_+   \cr
 &  - i    \psi_- (\partial_-z^j\Gamma^j + \partial_- y^{j'} \Gamma^{j'})
 ( z^i \Gamma^i - y^{i'} \Gamma^{i'} ) \psi_-
  \cr
& + \left. { 1 \over 12}  \partial_-(\tau + \chi) \left[ \psi_- \Gamma^{ij}
 \psi_ - \psi_- \Gamma^{ij} \psi_-
- \psi_- \Gamma^{i'j'} \psi_-  \psi_- \Gamma^{i'j'} \psi_- \right] \right\}\ ,
}}
where $\psi_\pm$ are real, positive-chirality $SO(8)$ spinors, $\Gamma^i$ are real $SO(8)$ gamma
matrices, and $\Pi \equiv \Gamma^1 \Gamma^2 \Gamma^3 \Gamma^4 $ is the product of the first four
gamma matrices.\foot{We can redefine $\psi_+ \to \Pi \psi_+$ to get rid of this matrix, which only
appears in the fermion mass term.}  The indices $i,~j$ run over the four transverse
directions in $AdS_5$, and the indices $i',~j'$ run over the four transverse directions in $S^5$.

As in the $O(N)$ case, we have the following chiral conserved currents
\eqn\conscur{\eqalign{
 j_+^\chi =& \partial_+ \chi - {y^2 \over 2 } +  i \psi_-\partial_+ \psi_-  + 2 i \psi_-
\Pi \psi_+ + {\rm 4 ~ fermi} ~,~~~~~~\partial_-j_+^\chi =0\ ,
\cr
 j_+^\tau = & \partial_+ \tau  + { z^2 \over 2 } -  i \psi_-\partial_+ \psi_-  - 2 i \psi_-
\Pi \psi_+ - {\rm 4 ~ fermi } ~,~~~~~~\partial_-j_+^\tau =0\ ,
 }}
 where we have not explicitly recorded the four-fermion terms proportional to the last term in
 \finlagrag . These currents are conserved due to the equations of motion for $\tau$ and $\chi$.
In addition, we have the conserved stress tensor
\eqn\stresscon{
T_{--} =  -(\partial_- \tau)^2 + ( \partial_- \chi)^2 +
(\partial_- z)^2 + (\partial_- y)^2  + 2 i (\partial_-\tau + \partial_- \chi)
\psi_- \partial_- \psi_-\ ,
}
obeying $\partial_+ T_{--}=0$.
So, as above, we have a field theory that is conformal for the right movers.
In this case, we expect that the field theory remains right-conformal, even after we include
quantum corrections.

Following the example above, we can now gauge-fix by imposing the conditions
\eqn\vircond{\eqalign{
 0=&j_+^\chi + j_+^\tau = \partial_+( \tau + \chi) + {z^2 -  y^2 \over 2 }\ ,
 \cr
 0=& T_{--}\ .
 }}
 We choose the coordinates\foot{These coordinates are similar to the ones chosen in \frolovlc .}
 \eqn\coordchoice{\eqalign{
 x^+ \equiv & \sigma^+ ~,~~~~  x^- \equiv 2 (\tau + \chi)\ ,
 }}
 so that the derivatives become
 \eqn\deriv{
  \partial_{\sigma^+} =   \partial_{x^+} + (z^2 - y^2)\partial_{x^-} ~,~~~~~ \partial_{\sigma^-}
  = 2 \left[\partial_{\sigma^-}(\tau + \chi) \right]\partial_{x^-}\ .
  }
We thereby obtain the following gauge-fixed Lagrangian:
 \eqn\ianlag{  \eqalign{
{\cal L} = & 4 \left\{
  \partial_+ \vec z \partial_- \vec z +  \partial_+ \vec y \partial_- \vec y  -
  { 1\over 4} ({\vec y}^2 + {\vec z}^2)
   + ( {\vec y}^2 - {\vec z}^2)
    [ ( \partial_- {\vec z})^2  + ( \partial_- {\vec y})^2] \right.
\cr   &
+ i \psi_+  \partial_- \psi_+ + i
 \psi_- \partial_+ \psi_-  + i \psi_-
\Pi \psi_+
 + i ({\vec y}^2 - {\vec z}^2) \psi_- \partial_- \psi_-
\cr
 & - i   \psi_-  (\partial_-z^j\Gamma^j + \partial_- y^{j'} \Gamma^{j'})   ( z^i \Gamma^i - y^{i'} \Gamma^{i'} ) \psi_-
  \cr
& \left. + {  1  \over 24 } \left[ \psi_- \Gamma^{ij} \psi_ - \psi_- \Gamma^{ij} \psi_-
- \psi_- \Gamma^{i'j'} \psi_-  \psi_- \Gamma^{i'j'} \psi_- \right] \right\}\ ,
}}
where we have also performed a simple rescaling of $\psi_+$.

Even though this action looks complicated, it is much simpler than the full gauge-fixed
Lagrangian that was written in \frolovlc .  These theories ought to have the full extended
$SU(2|2)^2 \times  R^2$ symmetry algebra described in \beiserts ,  so it is useful to
study the form of this algebra in the limit we are considering.
 We start with the
 supersymmetry algebra in \beiserts, which can be written as
 \eqn\limmicom{\eqalign{
 \{ Q^i_-,Q^j_- \} =& \delta^{ij}   \kappa_-      ~,~~~~~~\kappa_\pm = \kappa^0 \mp \kappa_1\ ,
\cr
\{ Q^i_-,Q^j_+ \} =& \delta^{ij} \kappa_2 + SU(2) ~{\rm currents}\ ,
\cr
\{ Q^i_+,Q^j_+ \} =& \delta^{ij}   \kappa_+\ .
}}
  We can think of $\kappa^\mu$ as a
$2+1$ dimensional momentum \hm ,
$\kappa^0= \epsilon$ as the ordinary energy, and $\kappa_{1,2}$ are the
two central charges introduced in \beiserts .

For the short representation corresponding to the elementary magnon, these charges obey
\eqn\condit{ \kappa_+ \kappa_- - \kappa_2^2 = (\kappa^0)^2 - \kappa_1^2 - \kappa_2^2 =  1\ .
}
For a single particle with momentum $k$, one obtains the values
\eqn\valcentr{ \kappa^1 + i \kappa^2 = - i 2 g e^{ i p_l }  ( e^{ i p} - 1)\ ,
}
where $p_l$ is the sum of the momenta to the left of the excitation.  The appearance
of this phase factor reflects the Hopf algebra structure of the
problem \magnonhopf.\foot{The presence of this phase is easy to understand
by drawing the pictures described in \hm . In that work the magnons are
represented as line segments that join points on a circle, and the momentum
is the angle subtended by these points. The
segment itself can be viewed as the complex number $\kappa_1 + i \kappa_2$.
The fact that the phase
depends on the number of magnons to the left is then clear: the magnon in question has to
be positioned on the circle at a point where the previous magnon ends, and
thus its orientation depends on the total angle, or momentum, subtended by all the magnons
to its left.}
Notice that, according to \condit, the
physical energy does not depend on the overall phase in \valcentr, and it is therefore
well-defined for each individual magnon.  In the near-flat-space limit that  we
 are considering, we can rescale the charges as $Q_\pm \to g^{\pm 1/4 } \hat Q_\pm$
and introduce similarly rescaled quantities
$\hat \kappa_\pm $ (we do not need to rescale $\kappa_2$).
After these operations, the final algebra
takes the same form as in Eqn.~\limmicom , but expressed in terms of rescaled quantities.
 The expressions for the rescaled central charges in terms of the momentum are
\eqn\exprech{\eqalign{
\hat \kappa_- =& 4 k ~,~~~~
\hat \kappa_+ = { 1 \over 4 k} + { k^3 \over 4 } + k^2 k_l + k k_l^2\ ,
\cr
\kappa_2 = & k ( k + 2 k_l)\ .
}}
We see that the central charges acting on a single magnon state still depend
on the momentum of the magnons to their left ($k_l$),
so they retain their Hopf algebra character.

At first sight it is a bit surprising that the
energy, which is conjugate to $x^+$, can be negative (as seen for large $k$ in
\energs).   In a supersymmetric system we might have expected the energy to be
positive.  In fact, the right-hand side of the $\hat Q_+$
anticommutator is $\hat \kappa_+$, and is not the energy (or $ i \partial_+$).
Indeed, $\hat \kappa_+$ is always positive.

Notice that $k$ is proportional to $\hat \kappa_-$. In fact, the right moving supercharges
$\hat Q_-$ act in the ordinary way on the Lagrangian \ianlag .  One can write the
Lagrangian \ianlag\ in terms of $(0,2)$ superfields by realizing explicitly two of the 8
right-moving supercharges. The action of the $\hat Q_+$ supercharges will be more non-trivial.
 The authors of \frolovalgebra\ have
expanded the action around the plane-wave limit up to terms quartic in the fields, so a limit
of their computation should give the proper symmetries of \ianlag .  (See also the
discussion in \roibanklose\  for further details.)
 We have not done this analysis here, and
it would be interesting to check explicitly that this Lagrangian indeed admits
the full symmetry algebra.

%%%%%%%%%%%%%%%%%%%%%%%%%%%%%%%%%%%%%%%%%%%%%%%%%%%%%%%%%%%%%%%%%%%%%%%%%%%%%%%%
\subsec{The  $S$ matrix in the weakly coupled regions}
\noindent
%%%%%%%%%%%%%%%%%%%%%%%%%%%%%%%%%%%%%%%%%%%%%%%%%%%%%%%%%%%%%%%%%%%%%%%%%%%%%%%%
The above theory becomes simple when the momenta of the excitations are small or large.
For small momenta, the excitations are described by ordinary
weakly-coupled massive quanta.  In the limit of very small
 momenta, they are free and the $S$ matrix goes to the identity for $k_i \to 0$.
The leading correction away from this limit is given by the computation done in
\refs{\swansonetal,\roibanklose}~ (see also \staudachers).
 In fact, it is simply the high-energy limit of the result in \refs{\swansonetal,\roibanklose}.
More explicitly, their result for the $\sigma$ factor is
\eqn\swansig{
\sigma^2 = 1 + { i \over 2 g } \tilde p_1^2 \tilde p_2^2 \left[
{ \tilde p_2 \over  1 + \epsilon_2} - { \tilde p_1 \over 1 + \epsilon_1} \right] {
1 \over ( \epsilon_1 + 1 ) (\epsilon_2+1) - \tilde p_1 \tilde p_2 }\ ,
}
where $\tilde p_1$ are the momenta rescaled to the plane-wave region
\eqn\tildepde{
\tilde p = 2 g p ~,~~~~~~\epsilon = \sqrt{ 1 + \tilde p^2 }\ .
}
By taking the near-flat-space limit \xpmdefw\  we find
\eqn\lowen{
\sigma^2 = 1 - i \, 2 \, k_1 k_2 { k_1 - k_2 \over k_1 + k_2 }\ .
}
This result is valid for small $k$, where the correction is small. As we increase $k$ we
should also consider higher-order corrections.

Similarly, we can consider the large-$k$ region.  In this region the magnons can be approximated
by classical solitons, and their scattering amplitude is a simple limit of the one
computed in \hm\ for the full theory.\foot{ Note that $S_0^{\rm here} = ( S_0^{\hm} )^{-1} $.}
The limit produces the following scattering phase
\eqn\scattp{
\log \sigma^2 \sim - i (k_1^2 - k_2^2 ) \log \left( {k_1 - k_2 \over k_1 + k_2 } \right)\ .
}
This result is valid as long as the right-hand side is large, which occurs for large $k$.

 Both of the results in Eqns.~\lowen\ and \scattp\  are summarized by the AFS
 \afs\ phase factor, which in this limit can be written as
\eqn\chiafs{
\sigma^2  = {  ( w_1^- + w_2^+ )^2 \over  (w_1^+ + w_2^-)^2   }
\left( {
(w_1^+ + w_2^-)(w_1^- + w_2^+) \over (w_1^+ + w_2^+) (w_1^- + w_2^-)  } \right)^{ i 2 (u_1 - u_2) }\ .
}
 However, we should emphasize that this is {\it not} the leading-order answer
 in the region where $k \sim 1$ or $w^\pm \sim 1 $, despite the fact that $g$ is large.
  In fact, in this region all higher order corrections should
 be taken into account, as we show more explicitly in the next section.

% JM

 We digress briefly to discuss the branch cut present in the semiclassical
 amplitude \scattp\ at $k_1-k_2 =0$.
 In principle, the exact phase can have branch cuts,
 but we do not expect that the exact phase would have branch
 cuts traversing regions of physical (real) momenta, or at least momenta that are
 very close to the real axis. So, for the purpose of this discussion, let us assume that
 the function $S(z)$, where $z= k_1-k_2$, is meromorphic.
  We expect that, as in other integrable theories,
 the branch cuts would be replaced by poles.

\ifig\zerospoles{ A closely spaced sequence of zeros or poles
should replace the branch cuts in the semiclassical scattering
amplitude \scattp . We expect to have zeros in the upper half
plane and poles in the lower half plane.  }
{\epsfxsize3.2in\epsfbox{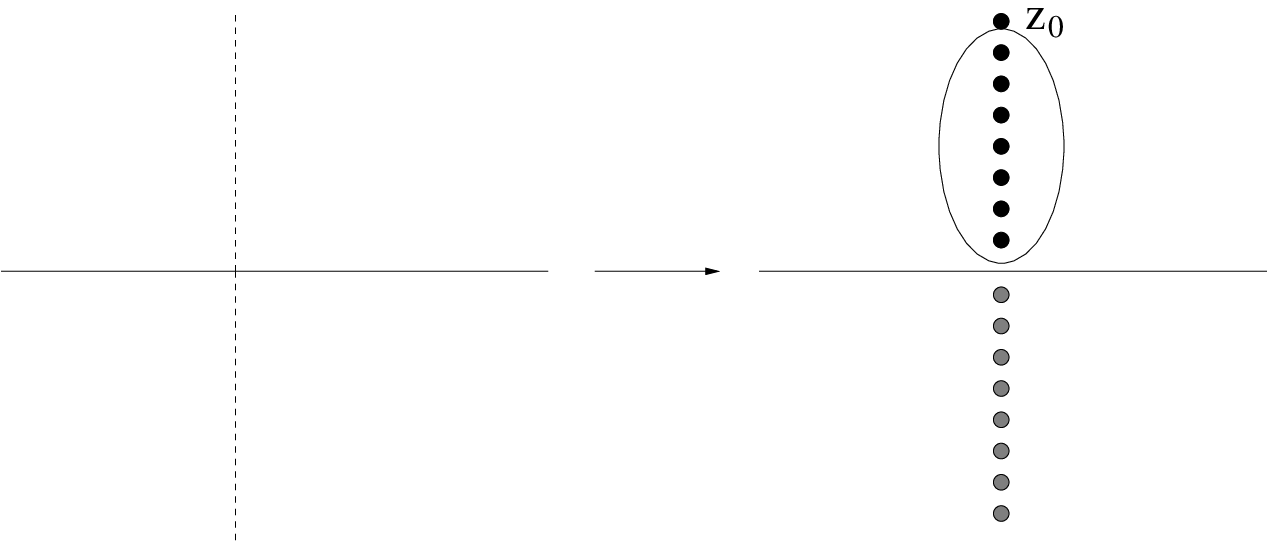}}

We know that the function $S$ has the approximate expression
 \eqn\forms{
 \log S \sim  h(z) \log z\ .
 }
 This expression should hold on the real axis and should fail when we
 approach the line of zeros or poles. Since the function should
 obey a unitarity condition $S(1,2) = 1/S(2,1)$, the existence of a
 pole in the upper half plane would imply a zero in the
 lower half plane. Let us assume, therefore, that there is single line of
 zeros in the upper half plane and a single line of poles in the
 lower half plane, both starting at $z=0$. Here, by a line of
 poles we mean a very closely spaced sequence. Thus, we will put
 half of the branch cut in \forms\ in the upper half plane and the
 other half in the lower half plane. As we cross the
 branch cut in the upper half plane the function jumps by $2\pi i h(z)$, and
 when we cross the lower one it jumps by $-i \pi h(-z)$. After we use
 that $h(z)=h(-z)$, we see that the total jump of the function as we go
 around a circle centered at $z=0$ is indeed $  2 \pi i h(z)$.
  (As an example, the reader could
 think about the function $S = {\Gamma( -i z) \over \Gamma(i z)}$.)
We now consider a contour that goes through $z=0$ and then circles
around the line of zeros in the upper half plane,
as depicted in \zerospoles . We then perform the following integral:
 \eqn\contint{
 N_p - N_z =  { 1 \over 2 \pi i } \oint d \log S
 \sim { 1 \over 2} h(z_0)\ ,
 }
 where  $N_{p,z}$ represents
 the number of poles or zeros enclosed by the contour.
 For this to make sense we need that $h(z_0)$ is real,
 which constrains the possible locations of zeros.
 Applied to the above problem, this logic would indicate the appearance of zeros in
 the upper half plane, where
 \eqn\polesget{
  N_z = - { i \over 2 }  (k_1^2 - k_2^2)\ ,
  }
  and $N_z$ is taken to be real and positive.
The formula \polesget\ should be interpreted as giving the mean
density of zeros, not their precise location.
  As discussed in \hm , the sigma model has localized time-dependent solutions,
  and one might be tempted to interpret these poles as physical bound states.
  Note, however, that this is not the
  only possible interpretation. In fact, if the poles turn out to be double poles,
  as we will see below
  (based on the  BES/BHL guess \refs{\bes,\bhl}), then another possible interpretation is that of
   ``anomalous thresholds''
  \coleman.\foot{This interpretation was suggested to us by N.~Dorey.} This issue
  will be addressed in a future publication.

%%%%%%%%%%%%%%%%%%%%%%%%%%%%%%%%%%%%%%%%%%%%%%%%%%%%%%%%%%%%%%%%%%%%%%%%%%%%%%%%
\newsec{Conjectured $S$ matrix in the near-flat-space limit}
\noindent
%%%%%%%%%%%%%%%%%%%%%%%%%%%%%%%%%%%%%%%%%%%%%%%%%%%%%%%%%%%%%%%%%%%%%%%%%%%%%%%%
As usual, we will parametrize the contribution to the scattering amplitude as
\eqn\contras{\eqalign{
\sigma^2 =&  e^{2  i \delta_{12}}\ ,
\cr
\delta_{12} = & \chi(w_1^+,w_2^+) -\chi(w_1^-,w_2^+)-
\chi(w_1^+,w_2^-)+ \chi(w_1^-,w_2^-) - ( 1 \leftrightarrow 2 )\ ,
\cr
\chi = & \sum_{n=0}^\infty   \chi^n\ ,
}}
where the $\chi^{n}$ represent contributions to the phase at $n$ loops.

%%%%%%%%%%%%%%%%%%%%%%%%%%%%%%%%%%%%%%%%%%%%%%%%%%%%%%%%%%%%%%%
\subsec{Tree-level and one-loop order}
\noindent
%%%%%%%%%%%%%%%%%%%%%%%%%%%%%%%%%%%%%%%%%%%%%%%%%%%%%%%%%%%%%%%
As mentioned above, the
 tree-level contribution in the scaling limit becomes \afs\
\eqn\afsans{
\chi^0(w_1,w_2) - \chi^0(w_2,w_1) = - (w_1^2 -w_2^2) \log(w_1 + w_2)\ .
}
The one-loop contribution turns out to be the simplest solution of the iterated  crossing
equation \crosseq . In fact, when written in the appropriate variables, the one-loop result in this
limit is the same as the one-loop result in the full theory.
We will begin by discussing a couple of aspects of the one-loop answer before taking the limit.

If one defines the variables \refs{\janikun,\bhl}
\eqn\thetavar{
x^\pm = \tanh { \theta^\pm \over 2 }\ ,
}
then the the double crossing\foot{As in \bhl , ``double crossing'' means that we iterate the
crossing transformation $x\to 1/x \to x$ along some particular path.} equation becomes
\eqn\doublcr{
\sigma^2_{\rm 1-loop}(\theta_1^+ + 2 i \pi, \theta^-_1 + 2 i \pi , \theta_2^+, \theta_2^- )
 =
{ \tanh^2 { \theta_1^+ - \theta_2^+ \over 2 } \tanh^2 { \theta_1^- - \theta_2^- \over 2 }
\over \tanh^2 { \theta_1^+ - \theta_2^- \over 2 } \tanh^2 { \theta_1^- - \theta_2^+ \over 2 } }
  \sigma^2_{\rm 1-loop}(\theta_1^+  , \theta^-_1   , \theta_2^+, \theta_2^- )\ ,
}
where $\sigma^2_{\rm 1-loop}$ is the one-loop contribution to the phase factor.
The simplest solution to this equation is \janikun
\eqn\simplestsol{
  \sigma^2_{\rm 1-loop} =  { h( \theta^+_1 - \theta_2^+ ) h(\theta_1^- - \theta_2^-) \over
h(\theta_1^+ - \theta_2^-) h(\theta_1^- - \theta_2^+ )  }\ ,
}
where $h$ is a function of the form
\eqn\hgunf{\eqalign{
h(\theta) = &  \prod_{n=-\infty}^\infty \left({  ( \theta - 2 \pi i  n)^2 \over
( \theta - 2 \pi i  (n  + {1 \over 2}) )( \theta - 2 \pi i   (n  - {1 \over 2}) )} \right)^n\ ,
\cr
h(-\theta) = & 1/h(\theta) ~,~~~~~~~~~~~h(\theta + 2 \pi i ) = \tanh^2 { \theta \over 2 } h(\theta)\ .}
}
As shown in \bhl, this solution is actually the same as the one-loop contribution to the
$S$ matrix \hl .

Now, as we take the near-flat-space limit,  we can expand
\eqn\newlim{\eqalign{
x^\pm =& \tanh { \theta^\pm \over 2} \sim 1 - 2 e^{ - \theta^\pm }\ ,
\cr
\hat \theta^\pm  =& \theta^\pm  - \log ( 2  \sqrt{g})   + i \pi\ ,
\cr
w^\pm = &  e^{ - \hat \theta^\pm }\ .}
}
Note that certain quantities such as
$\theta_1^\pm - \theta_2^\pm = \hat \theta^\pm_1 - \hat \theta^\pm_2 $ remain the same in
the limit.
Thus the one-loop answer is the same as \simplestsol, except that we replace
$\theta^\pm  \to \hat \theta^\pm $, which is related to $w^\pm$ through \newlim .

%%%%%%%%%%%%%%%%%%%%%%%%%%%%%%%%%%%%%%%%%%%%%%%%%%%%%%%%%%%%%%%%%%%%%
\subsec{Higher orders}
\noindent
%%%%%%%%%%%%%%%%%%%%%%%%%%%%%%%%%%%%%%%%%%%%%%%%%%%%%%%%%%%%%%%%%%%%%
In BES/BHL \refs{\bes,\bhl},
 a concrete proposal for the $S$ matrix was made. The answer was expressed as a series
expansion in $1/g$ and $1/x_1,~1/x_2$.  Since we are interested in the region near $x_1 \sim 1$,
we will need to sum the series expansion in $1/x$. After this is done
(for details, see Appendix D) we
obtain
\eqn\chiexpr{\eqalign{
\tilde \chi^n = &- {\zeta(n) \over (-2 \pi)^n }{ 1 \over x_1 x_2^2} { \Gamma(n-1) \Gamma({n\over 2} ) }
\left[ g ( 1 - { 1 \over x_2^2})(1 - { 1 \over x_1x_2}) \right]^{1-n}
\cr
& \times \sum_{l=0}^{n-2} \sum_{q=0}^l
\sum_{m=0}^{n-2} (-1)^{q+m}
 {\Gamma( { n\over 2} + l - q)  \over q!  \Gamma(1 + l -q) \Gamma({ n \over 2} -q) }
\cr
  &  \times  {  \Gamma(n-1 + l -q  - m)
 \over \Gamma(  { n \over 2}  + l - q - m) m! \Gamma(n-1 -m)}
 \left({ 1 - { 1 \over x_1x_2} \over{  1- { 1 \over x_2^2}}} \right)^l
  (1-{1 \over x_2^2})^{m +q }\ .}
}
These expressions manifestly display the singularities at $x_i =1$, and are valid for $n\ge 2$.

We can now take our near-flat-space limit
 in \xpmdefw, and we find that only the leading singular terms with
$q=w=0$ in \chiexpr\ contribute:
\eqn\fullnoth{\eqalign{
 \chi^n(w_1,w_2) \equiv & \lim_{g\to \infty} \tilde \chi^n(x_1,x_2)\ ,
\cr
  \chi^n(w_1,w_2) \ = &  - { \zeta(n) \over (-2 \pi)^n } { 1 \over (w_1 + w_2)^{n-1} (2 w_2)^{ n -1} }
\sum_{l=0}^{ n -2} {\Gamma( n + l - 1) \over \Gamma( l +1) }{ (w_1+ w_2)^{l  }  \over (2w_2)^l
 }\ .}
}
As anticipated, we need to keep all orders in $\alpha'$ in this regime. The weak
coupling expansion corresponds to the expansion in powers of $1/w$.
The expressions in Eqns.~\chiexpr\ or  \fullnoth\
are such that if $x_1 \sim 1$, but $x_2 \sim -1$, then there
is no contribution as $g \to \infty$.
Similarly, the one loop contribution also vanishes in
this limit as $\theta^\pm_1 - \theta^\pm _2 \to \infty $ in \simplestsol, and we use that
$h \to e^{-i \pi/4}$ in that limit, so that $\sigma^2_{\rm 1-loop} \to 1$.
Then in the region
$x_1 \sim 1$ and $x_2 \sim -1$, the leading contribution is just the tree-level contribution
\afsans, which was shown in \afs\ to reproduce precisely the flat-space spectrum.
In other words, the BES/BHL guess \bes\ has a structure that renders the computation of the
flat-space spectrum in \afs\ valid.\foot{This of course holds with the caveat we mentioned
above regarding the need to take $J \sim \sqrt{g}$, which might
make the asymptotic analysis unreliable.}

One nice aspect of this strong coupling series is that it is Borel summable.
As shown in detail in
Appendix $D$, we can write the full sum over $n$ as
\eqn\fullch{\eqalign{
 \hat \chi(w_1,w_2) = &
\sum_{n=2}^\infty  \chi^n(w_1,w_2)\ ,
\cr
\hat \chi(w_1,w_2) = &    { 1 \over 2 \pi  } \int_0^\infty d\tau   {1  \over w_1
 + w_2  +  \tau }
    \log \left[ 1 - e^{ -  2 \pi  \tau^{ 2 } - 2 \pi (2 w_2) \tau    } \right] \ .}
}
This integral defines the sum exactly for all values of $w^\pm$.
Note that the odd-$n$ contributions were essential to be able to Borel sum the
expression.\foot{Another double-scaling limit was considered in \gh , where only even $n$
were summed, which led to singularities.}  The full phase factor is then
$\chi = \chi^0 + \chi^1 + \hat \chi$.

\ifig\contours{ Contours that we should choose for performing
the crossing
transformation. We need to enforce the the constraint
 $(w^+)^2 -(w^-)^2 = i$, and move $w^+$ and $w^-$ together as we
 do the crossing transformation.
 The dots sit at $\pm e^{\pm i \pi/4}$. }
{\epsfxsize2.5in\epsfbox{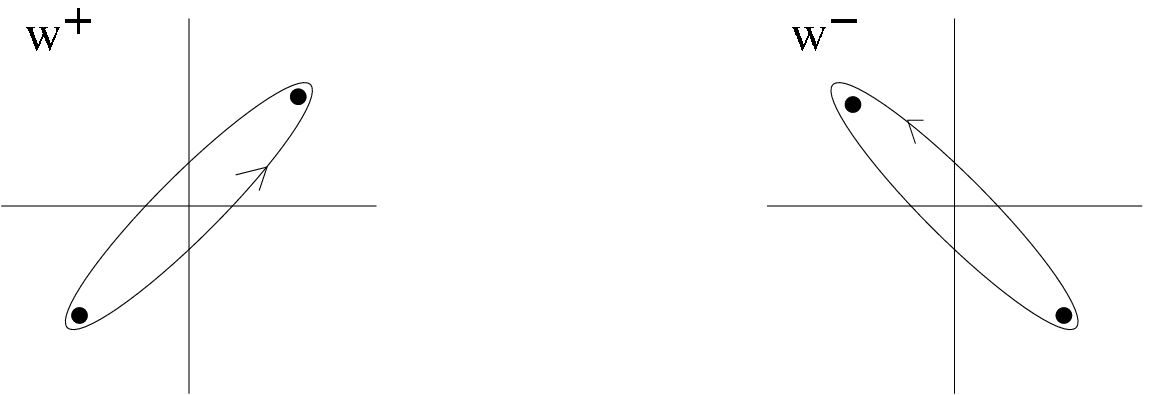}}

It is also possible to write a combined integral expression for $\chi$ of the form
\eqn\finantisym{
\chi(w_1,w_2) - \chi(w_2,w_1) = { 1 \over 2 \pi }
 \int^{+\infty}_0 { d z\over w_1 + w_ 2 + z }  \log \left\{ { \sinh[ \pi z
 (z + 2 w_2)]  \over \sinh [ \pi z
 (z + 2 w_1) ] } { (z + 2 w_1 ) \over (z + 2 w_2 ) }  \right\}\ .
}

One can perform a few checks on this expression.
First, one would like to check that the expression obeys the crossing symmetry
equation. All the contributions from loop orders two and beyond, constructed from
\fullch,  should not change under double crossing, but should change in
a very specific way under single crossing \bhl .
We can see that this is indeed the case, provided we choose the crossing contour shown
in \contours .  We can characterize the contour
in terms of the variable $\eta$ introduced in
\wpmfrometa\ by stipulating that $\eta = \epsilon + i t$, with
$\epsilon$ small and positive. The crossing transformation corresponds to $t \to t + \pi$,
while double crossing takes $t \to t + 2 \pi i$.  Of course, the starting point for the crossing
transformation need not be on this contour, in which case we deform the contour in such a
way that we go around any possible branch cuts.
This choice of contours is important because the integrand has singularities at
\eqn\singint{
\tau = - w_2 \pm \sqrt{ w_2^2 + i n }\ ,
}
for all integers $n\in {\bf Z}$, and these might lead to additional contributions.
 These points lead to branch cuts in $\hat \chi$ at $w_2 = \pm \sqrt{i n}$.
 These branch cuts were not present in each of
the individual terms in \chiexpr\  and, for this reason, it is nontrivial
that there exists a contour that allows the crossing transformation to work.
The explicit check of the crossing equation is discussed in more detail in Appendix D.

Note that the final result is a smooth function without singularities as long as
$Re(w_1),~ Re(w_2)>0$.
This condition is obeyed at the physical (real) values of the momenta \constrbe .
Thus, we  smoothly interpolate between the small $k$ and large $k$ regimes.
At this point it should be checked that the large-$k$ answer is
indeed the one we expect from Eqn.~\scattp , and we want to see how the
branch cut in \scattp\ is replaced by poles.
At first sight we seem to have a problem, since the leading order result \chiafs\
has a branch cut at $w_1 + w_2 \sim 0$.
In fact, Eqn.~\fullch\  also contains a branch cut that precisely cancels the leading-order
result \chiafs .  This branch cut arises from the pole of Eqn.~\fullch\ at $\tau = -w_1 -w_2$.
This cancellation is most easily seen by looking at \finantisym , which is manifestly
non-singular
at $\tau = - w_1 - w_2 $.

In addition, we can see that we will get additional singularities whenever two singularities
of the integrand, which are at $\tau = - w_2 + \sqrt{ w_2^2 + i n}$, pinch the integration
contour.
All these singularities arise away from the real physical values of momenta.  However, for
large momenta they can lie very close the the physical values.
 Focusing on the singularities that are very close to the physical subspace, we can
show that the full factor $\sigma^2$ contains double poles at
\eqn\doublepo{ \sqrt{(w_2^+)^2 + i n} + w_1^- =0\ , } where $n>0$.
Appendix D gives the derivation of this result. Here we define the
branch of the square root in such a way that its real part is
positive when the real part of $w_2^+$ is positive. Of course, we
also have  double zeroes at \eqn\doublezer{ \sqrt{(w_1^+)^2 + i n
} + w_2^- =0\ . }
We also find that the one loop expression has a single zero at $w_2^+ + w_1=0$, which
is cancelled by a pole in the sum of all the higher-order terms. The net result is that the
$\sigma^2$ factor does not have a zero or pole at $w_2^+ + w_1^-=0$ (see appendix D).
In fact, it is analytic
at this point. This is again most easily seen by looking at \finantisym . \foot{The
first version of this paper had an incorrect statement on this point.}
%extra factor in \crosseq, we find that $S_0$ also contains a double pole at
%$w_2^+ + w_1^-=0$, and a corresponding double zero at $w_1^+ + w_2^-=0$.
%We should therefore include the case with $n=0$ in Eqns.~\doublepo , \doublezer .
%(The reader is referred to Appendix D for further details).
%One of the guesses in \bhl, dubbed ``giant'', is a different function but
%also exhibits these poles.\foot{For example, we will see that the function $\hat\chi $
%discussed here has a branch cut at $w_i^\pm = 0$, but the ``giant'' guess in \bhl\
%does not.}

\ifig\poles{  We fix a real value for $u_2$ and
 display the poles and zeros
 in the complex $u_1$ plane.  There is a branch cut in the amplitude starting
 at $u_1 = \pm i/2$, and possible additional branch cuts appear at larger imaginary values.
If $u_2 <0$, as in (a),
 the poles are immediately accessible by moving $u_1$ in the complex direction. They are denoted
 by dots. If
 we start with $u_2 >0$, as in (b), the poles are in a second branch that is accessible only
 after taking $u_1$ through the branch cut. We denoted these by crosses.  Thus, for $u_2>0$ we do not
 encounter poles when we move $u_1$ in the imaginary direction.   }
{\epsfxsize2.5in\epsfbox{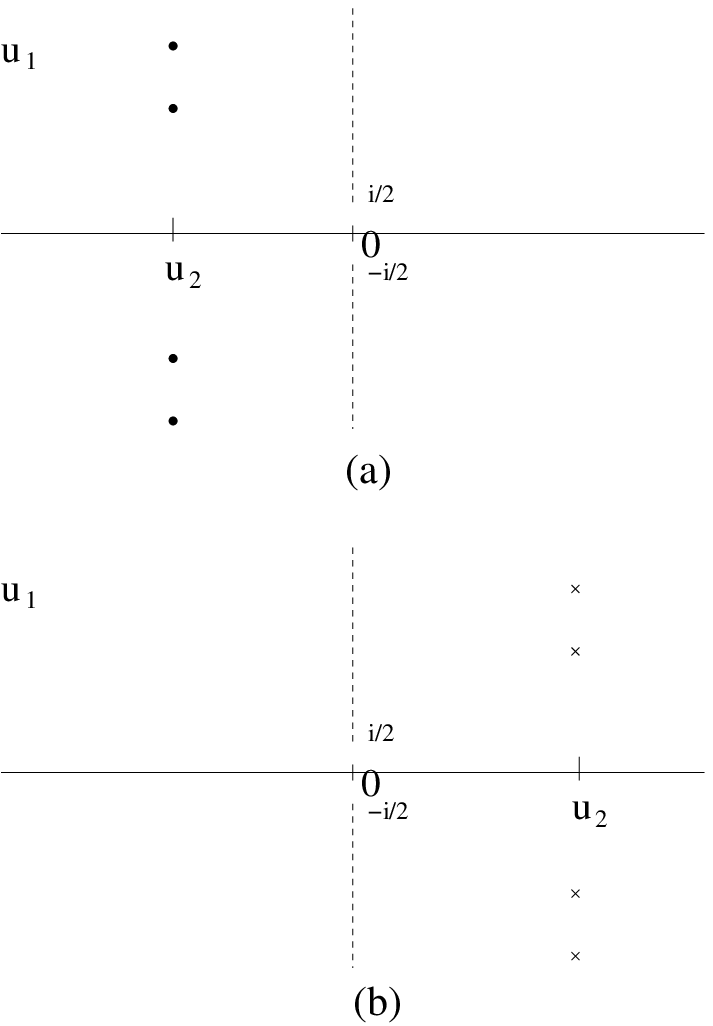}}

We can also express Eqn.~\doublepo\ as
\eqn\condie{
u_1 - u_2 = i n ~,~~~~~n>1\ ,
}
but here we are losing information since we do not distinguish between the two possible
signs in the square root on Eqn.~\doublepo.  In fact, we do not find these poles if
we start with large positive $u_1 \sim u_2$ and analytically continue in the imaginary
direction.  We only find them if we start with {\it negative} $u_i$ and analytically continue.
This structure is shown in  \poles .  We get a mean density of poles consistent
with the one sketched at the end of Section 3.3.\foot{Note
that $ u_1 - u_2 \sim - { 1 \over 4 } ( k_1^2 - k_2^2 ) $
for large $k_i \gg 1 $. }

Note that it is a nontrivial check of BES/BHL \refs{\bes,\bhl} that we get
the expected results in the giant-magnon region. In fact, some of the other
guesses in \bhl\ do not have this property.  Note, in particular, that the elementary
excitations and the giant magnon only differ by their momentum, and they are
continuously connected and do not constitute two different kinds of excitations.

%%%%%%%%%%%%%%%%%%%%%%%%%%%%%%%%%%%%%%%%%%%%%%%%%%%%%%%%%%%%%%%%%%%%%%%%%%%%%%%%%
\newsec{Conclusions}
\noindent
%%%%%%%%%%%%%%%%%%%%%%%%%%%%%%%%%%%%%%%%%%%%%%%%%%%%%%%%%%%%%%%%%%%%%%%%%%%%%%%%%
In this article we have studied some aspects of an especially
 interesting limit of the worldsheet $S$ matrix of type IIB string theory
on $AdS_5 \times S^5$.  Since the limit is one in which the spacetime curvature radius
is taken to infinity, one might have
thought that we would arrive at a free theory.  Nevertheless, the effects of the leading
deviation away from flat space are still important in this limit.  The reason is that
the excitations have a long time to interact:
the effects of small curvature are thus compounded, producing large contributions.

In this limit, the theory
 has many of the features that are encoded in the full $S$ matrix, such as
the fact that the magnon-scattering is off-diagonal and admits a non-trivial crossing-symmetry
equation.   Moreover, we have found that the Lagrangian in this limit looks
fairly simple, and might be exactly solvable in some fashion.
Since the string theory ultimately becomes weakly coupled, one would expect that
there is a clever way to solve this model directly.

We also studied the BES/BHL \refs{\bes,\bhl} proposal for the $S$
matrix and found that it correctly reproduces the salient
properties of the giant-magnon region, at least in this limit. We
also found that we have double poles in this region. The physical
interpretation of these double poles will be addressed in a future
publication.

~~~~~~~~~~~~~~~~~~~~~~~~~~~~~~~~~~~~~~~~~~~~~~~~~~~~~~~~~~~~~~~~~~~~~~~~~~~~~~~~~~~~~~~~~

\noindent
{\bf Acknowledgments }

\noindent
We thank A.~Neitzke for collaboration at an early stage in this project.
We thank N.~Beisert, N.~Dorey, S.~Hellerman, D.~Hofman, R.~Janik,
T.~McLoughlin and R.~Roiban
for discussions.  We also thank T.~McLoughlin and J.~Minahan
for pointing out sign and coefficient errors in an earlier draft.
I.S.~is the Marvin L.~Goldberger Member
at the Institute for Advanced Study, and is supported additionally
by U.S.~National Science Foundation grant PHY-0503584.
The work of J.M. was  supported in part by U.S.~Department of Energy
grant \#DE-FG02-90ER40542.

%%%%%%%%%%%%%%%%%%%%%%%%%%%%%%%%%%%%%%%%%%%%%%%%%%%%%%%%%%%%%%%%%%%%%%%%%%%%
\appendix{A}{ Analytic properties implied by the crossing relation}
\noindent
%%%%%%%%%%%%%%%%%%%%%%%%%%%%%%%%%%%%%%%%%%%%%%%%%%%%%%%%%%%%%%%%%%%%%%%%%%%%
We introduce the variable $\eta$ defined via
\eqn\devarn{
w^+ = i^{1/4} \cosh \eta ~,~~~~~~~~w^- = i^{1/4} \sinh \eta\ .
}
Crossing symmetry shifts $\eta \to \eta + i \pi $.
Let us write the crossing symmetry equation \crosseq\ \janikeq\ in terms of these variables.
We find
\eqn\janebqu{\eqalign{
 \sigma^2(\eta_1 + i \pi , \eta_2) \sigma^2(\eta_1,\eta_2) =&
 {
 ( \sinh \eta_1 - \cosh \eta_2 )^2 \over (\cosh \eta_1 + \sinh \eta_2 )^2}
 {
 (\sinh \eta_1 + \sinh \eta_2 )^2 \over ( \cosh \eta_1 - \cosh \eta_2 )^2 }
 \cr
=&  {
 ( \sinh \eta_1 - \cosh \eta_2 )^2 \over  (\cosh \eta_1 + \sinh \eta_2 )^2 } { 1 \over
  \tanh^2 {\eta_1 -\eta_2 \over 2 } } \ , }
}
 where we have simplified the second factor.  We can iterate the equation once to
 obtain the equation
\eqn\janecqu{\eqalign{
  \sigma^2(\eta_1 , & \eta_2 - 2 \pi i ) =\sigma^2(\eta_1 +2 i \pi , \eta_2)  =
  \cr
  & \sigma^2(\eta_1,\eta_2) {
   (\cosh \eta_1 + \sinh \eta_2 )^2  \over( \sinh \eta_1 - \cosh \eta_2 )^2   }
  {
 ( \sinh \eta_1 + \cosh \eta_2 )^2 \over (-\cosh \eta_1 + \sinh \eta_2 )^2  }
  \tanh^{4}{\eta_1 -\eta_2 \over 2 }\ ,
}
}
where we have used the equations for crossing of $\eta_1$ and $\eta_2$.
We see that $\sigma^2$ should pick up the above factor when we shift its argument
by $2 \pi i$. If we had just the last factor involving the hyperbolic tangent
it would be easy to find a meromorphic solution to \janecqu .\foot{ The solution is just $h^2$, with
$h$ in \hgunf.}
The problem arises from the first factor.
We will now show that there exists no meromorphic solution, as a function of
$\eta_1$ and $\eta_2$, which solves \janecqu .

\ifig\etaonetwo{  Contour in the $\eta_1$ plane that leads to a shift in
$\eta_2$ when we impose the equation $\sinh \eta_1 - \cosh \eta_2
=0$. Letters indicate points that are mapped to each other,
  and the contour on the left maps into the contour on the right.   }
{\epsfxsize4.0in\epsfbox{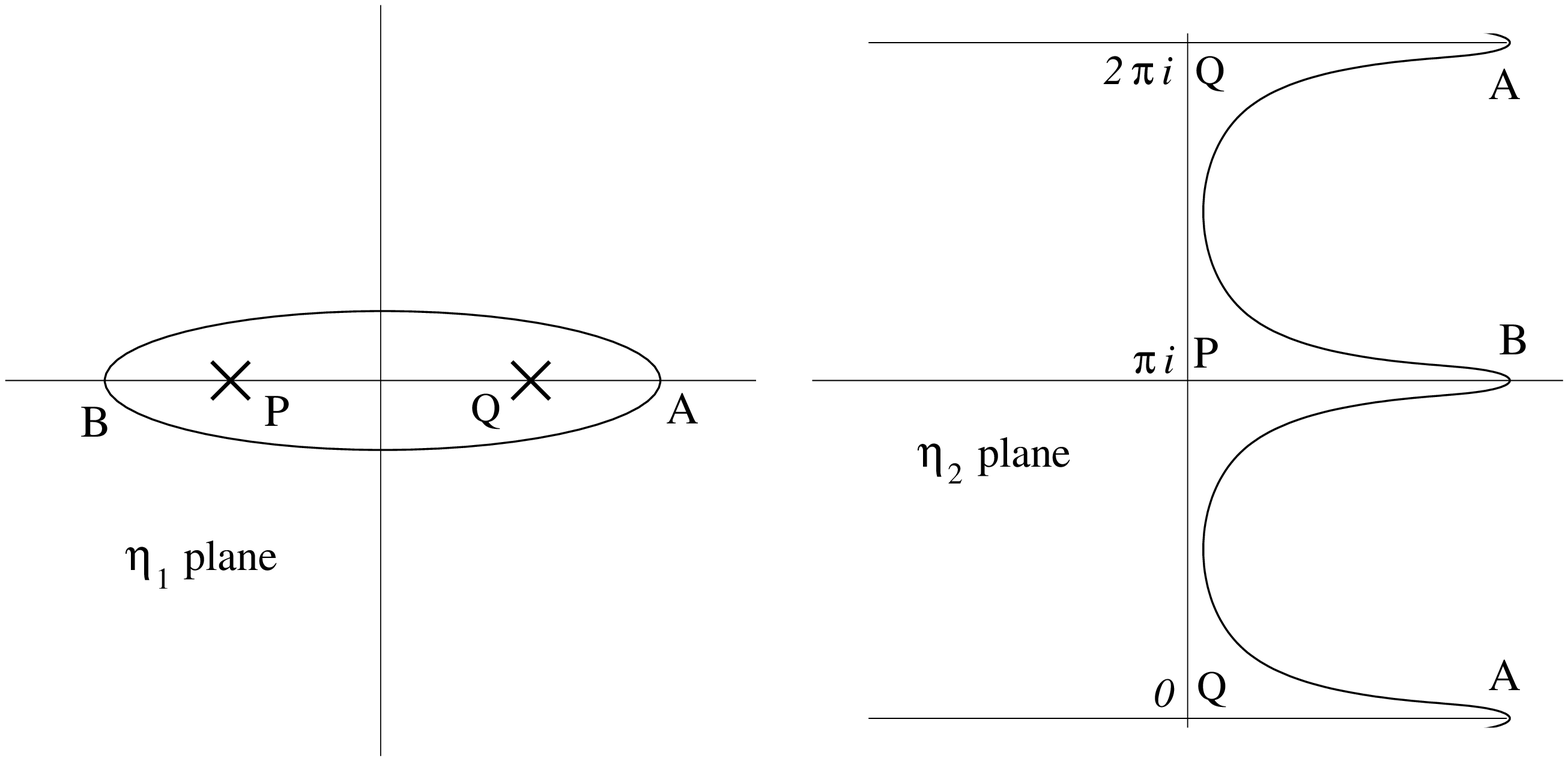}}

First, we focus on a particular pole developed by one of the factors
in \janecqu :
 \eqn\zerocond{
\sinh\eta_1 - \cosh \eta_2 =0\ .
}
Solving for $\eta_2$ as a
function of $\eta_1$, we see that we have ${\bf Z}_2$ branch points
where $\sinh \eta_1 = \pm 1$. If we take a contour in the complex $\eta_1$ plane
going around these two branch points, we see that
$\eta_2(\eta_1) $ will change by $2 \pi i $.  This contour is depicted in \etaonetwo .

Now, let us imagine that we obtain a putative meromorphic
solution $\sigma^2(\eta_1,\eta_2)$.   Consider the holomorphic
function $f(\eta_1) = \sigma^2(\eta_1,\eta_2(\eta_1))$, and
assume that for some value of $\eta_1$ the function $f$ is neither zero nor infinite.
Then, by assumption, we will get a meromorphic function $f(\eta_1)$,
which could have branch points where
$\eta_2(\eta_1)$ exhibits branch points.  However, by continuing $\eta_1$ along
the contour in \etaonetwo , so that $\eta_2(\eta_1) \to \eta_2(\eta_1) + 2 \pi i$,
we find that $f(\eta_1)$ becomes identically zero by virtue of the crossing relations.
This is a contradiction, because we would find that $f$ becomes zero
after going through a branch cut.
If $\sigma^2(\eta_1,\eta_2)$ has a finite-order pole or zero at $\eta_2 = \eta_2(\eta_1)$, then
we can repeat this argument after we go around the contour a sufficient number of times so as
to cancel the given pole or zero.

To evade this contradiction we need to either allow branch cuts in $\sigma^2$ or allow essential
singularities along the regions where \zerocond\ is obeyed.\foot{Solutions with
essential singularities do exist \janikun.}
The option with essential singularities appears to be incorrect, since the
line of essential singularities approaches the real physical line,
for example, as $w^-_1= w_2^+ \to \infty $.

Even though we have made this argument in this special near-flat-space limit,
the same argument can be made for the full crossing-symmetry equation in \janikeq .
In this case one may start with the assumption of a meromorphic solution
on the two planes that result from going to
the cover of the two tori appearing in the description used in \janikeq .

%%%%%%%%%%%%%%%%%%%%%%%%%%%%%%%%%%%%%%%%%%%%%%%%%%%%%%%%%%%%%%%%%%%%%%%%%%%%%%%%%%%%%%%%%%%
\appendix{B}{Some properties of the $O(N)$ sigma model in the near-flat-space limit}
\subsec{Classical integrability}
\noindent
%%%%%%%%%%%%%%%%%%%%%%%%%%%%%%%%%%%%%%%%%%%%%%%%%%%%%%%%%%%%%%%%%%%%%%%%%%%%%%%%%%%%%%%%%%%
Consider the theory
\eqn\theorcov{
S = \int \partial_+ \vec y \partial_-\vec y + \partial_+ \chi \partial_-\chi - \vec y^{\, 2} \partial_-
\chi\ .
}
The Lax connection can be obtained by taking a simple limit of the connection
for the $O(N)$ theory \sosix .
To write it down explicitly, let us select one of the $O(N)$ generators, $J_{12}$, and consider
the off-diagonal generators that mix the $(1,2)$ plane with the rest:   we call them
$J^{\pm i}$, where $i=1,\cdots , N-2$.  We will need the following commutation relations:
\eqn\generat{\eqalign{
[ J^{12} , J^{\pm i} ] &= \pm J^{\pm i }\ ,
\cr
 [J^{+ i},J^{- j}] &= - \delta_{ij} J^{12} - J^{ij} ~,~~~~~~~~~
[J^{-  i},J^{ + j}] =  \delta_{ij} J^{12} - J^{ij}\ .
}}
We then construct the flat connection ${\cal A}(w)$, where $w$ is the spectral parameter
\eqn\laxcon{\eqalign{
 d + {\cal A} =& d + { \cal A}_+ d \sigma^+ + { \cal A}_- d \sigma^- ~,~~~~~~~~d {\cal A} + {\cal A}^2  =0\ ,
\cr
{ \cal A}_+ &={ i \over \sqrt{2}} \left[
 e^{ - i \sigma^+ w}   y^i J^{+ i} +
 e^{ + i \sigma^+ w}   y^i J^{- i} \right]\ ,
\cr
{\cal A}_- & =  { 1 \over w} \left[ - i \partial_- \chi J^{12}  - { 1 \over \sqrt{2}}  e^{ - i \sigma^+ w}
\partial_- y^i J^{+ i} + {1 \over \sqrt{2}} e^{  i \sigma^+ w} \partial_- y^i J^{- i }  \right]\ .
}}

We can now use the new variables \newcor\ and the gauge-fixing conditions $j_+=T_{--}=0$ to
derive the Lax connection for the gauge fixed Lagrangian \intlagr . It will be convenient
to employ the relations
\eqn\relsc{
d\sigma^+ = dx^+ ~,~~~~~~d\sigma^-2({1\over 4}  + \partial_{\sigma^-} \chi) = dx^- - y^2  dx^+\ .
}
Using these equations and the constraints we find the new connection
\eqn\newconn{\eqalign{
& d + { \cal A}_+ dx^+ + {\tilde {\cal A}} (dx^- - { y^2 } dx^+)\ ,
\cr
{\tilde {\cal A}} &\equiv { 1 \over w} \left[ -i   ( {1 \over 4}  - ( \partial_{x^-} y)^2 ) J^{12}
 - { 1 \over \sqrt{2}}  e^{ - i x^+ w}
\partial_{x^-} y^i J^{+ i}
%+\right.
%\cr
% & \left.
+ { 1 \over \sqrt{2}} e^{  i x^+ w} \partial_{x^-} y^i J^{- i }  \right]\ ,
}}
and $ {\cal A}_+$ is the same as what we had above in \laxcon.\foot{Note that the
``$+$'' component of the connection in these variables is
${\cal A}_+ - y^2 \tilde{\cal A} $. }

One can perform a gauge transformation to remove the constant part of the connection
\eqn\gaugetr{
 {\cal A} \to { \cal A}' =  g^{-1}  {\cal A} g + g^{-1} d g ~,~~~~~~g = e^{ i { 1 \over 4 w } x^ -
 J^{12} }\ .
}
We then find that the quantities in the above equation become
\eqn\newcon{\eqalign{
 & d + { \cal A}'_+ dx^+ + {\tilde {\cal A}}' (dx^- - { y^2 } dx^+)\ ,
\cr
{ \cal A}'_+ &={ i \over \sqrt{2}} \left[
 e^{ - i x^+ w - i { 1 \over 4 w } x^-}   y^i J^{+ i} +
 e^{   i x^+ w + i { 1 \over 4 w } x^- }   y^i J^{-i} \right]\ ,
\cr
{ \tilde {\cal A}}' &= { 1 \over w } \left[ i     ( \partial_{x^-} y)^2   J^{12}
 - { 1 \over \sqrt{2}}  e^{ - i x^+w + i { 1 \over 4 w } x^-}
\partial_{x^-} y^i J^{+i} +\right.
\cr
& \left. + { 1 \over \sqrt{2}} e^{  i x^+ w + i { 1 \over 4 w }
 x^-} \partial_{x^-} y^i J^{-i }  \right]\ .
}}

%%%%%%%%%%%%%%%%%%%%%%%%%%%%%%%%%%%%%%%%%%%%%%%%%%%%%%%%%%%%%%%%%
\subsec{The quantum theory}
\noindent
%%%%%%%%%%%%%%%%%%%%%%%%%%%%%%%%%%%%%%%%%%%%%%%%%%%%%%%%%%%%%%%%%
We now consider the quantum theory based on \theorcov . To avoid IR problems,
it is convenient to expand the theory around a vacuum where $\partial_- \chi = m^2$.
In this way the field $y$ becomes massive.
We obtain a divergent diagram from the one-loop self energy of $y$
of the form
\eqn\divdia{
I(m^2) = (N-2) { 1 \over 2}  \int { d^2 p \over (2 \pi)^2 } \log( p^2 + m^2 )\ .
}
We can see that
\eqn\deri{
\partial_{m^2 }I = { (N-2) \over 2}
\int { d^2 p \over (2 \pi)^2 } {1 \over  p^2 + m^2 } \sim { (N-2) \over 8 \pi}
\log ( \Lambda^2/m^2)\ .
}
Thus it is clear that $I \sim m^2 \log m^2$.  This violates the right-conformal
symmetry $\sigma^- \to \lambda \sigma^- $ that rescales $m^2$.
This is perhaps not surprising, given that the original $O(N)$ sigma model is not conformal once
quantum corrections are taken into account.
 The theory is therefore not right-conformal in general.

Even though the quantum theory is not conformal
 it is still integrable, and one can solve it using the known solution for the $S$ matrix for
 the $O(N)$ sigma model \zamolodchikov .
  Note that,
for the particular case of $N=2$, the theory is conformal. At first sight this theory
is trivial, since in that case we do not have any $y$. However, as stressed in \mannpolchtwo ,
we can view this theory as one based on the $OSp(M+2|M)$ supergroup. In this case
we would have $M$ bosonic $y$ fields and $M$ fermionic $y$ fields.
The $OSp(M+2|M)$ theory was studied
in \refs{\mannpolchone,\mannpolchtwo} as a toy model for the full $AdS_5 \times S^5$ sigma model.
One can similarly study it in the limit we consider here, where some of the equations in \mannpolchtwo\
simplify a bit. Using those results, one can show that the scattering of impurities in this model
gives rise to double poles.

%%%%%%%%%%%%%%%%%%%%%%%%%%%%%%%%%%%%%%%%%%%%%%%%%%%%%%%%%%%%%%%%%%%%%%%
\appendix{C}{A proof of formula \chiexpr }
\noindent
%%%%%%%%%%%%%%%%%%%%%%%%%%%%%%%%%%%%%%%%%%%%%%%%%%%%%%%%%%%%%%%%%%%%%%%
In this section we consider the sums that appear in the definition of the phase.
The starting expressions will be the conjectured forms in BES/BHL \refs{\bes,\bhl},
 for $n\geq 2 $:
\eqn\somedefi{\eqalign{
\chi^n =&  (-1) { \zeta(n) \over (-2 \pi)^n} {1 \over g^{n-1} \Gamma(n-1) } { x_1 \over (x_1 x_2)^2 }
\hat \chi ~;\cr
 \hat \chi &\equiv
\sum_{t=0}^\infty \sum_{m=0}^\infty { \Gamma( t+1 + m + { n \over 2 }) \Gamma( m + { n \over 2} )
\over \Gamma(t + 3 + m - { n \over 2} ) \Gamma( 2 + m - { n \over 2} ) } { 1 \over x^t y^m }\ ,
\cr
r_{_{\bhl}} & = t + 2 ~, ~~~~~s_{_{\bhl}} = r + 1 + 2 m  ~,~~~~~~~  x \equiv x_1 x_2 ~,~~~~~~~y \equiv x_2^2\ .
}}
In the last line we related the summation indices in \bhl\ to our own.
Using Mathematica, we can easily show that \somedefi\ is equal to \chiexpr, for low values of $n$.
These were also summed in \bhl .
One can give a general proof as follows.
When the sums are performed, one typically generates hypergeometric functions.  We will use
the following identity:
\eqn\identu{\eqalign{
{ 1 \over \Gamma(\tilde m + b - \tilde n) } F(\tilde m ,b, \tilde m+ b - \tilde n, z) =&
{\Gamma(1-\tilde m+\tilde n) \over \Gamma(\tilde m ) \Gamma(b-\tilde n) \Gamma(b) }
\cr
& \times \sum_{s=0}^{\tilde n-\tilde m}
 { \Gamma(b-\tilde n+ s)  \Gamma(\tilde n-s) \over s!
 \Gamma(1 -\tilde m + \tilde n -s ) } (1 - z)^{s -\tilde  n}\ ,
}}
which is true if $\tilde m, ~\tilde n$ are integers, and $0<\tilde m\leq \tilde n$.

The first step will be to do the sum over $t$ in Eqn.~\somedefi . This gives
\eqn\finsumt{
\sum_{t=0}^\infty { \Gamma(1 + t + m + n/2) \over \Gamma(t + 3+ m - n/2)} x^{-t} =
{\Gamma(n-1) \over \Gamma(2 + m - { n \over 2 } )} \sum_{s=0}^{n-2} { \Gamma(2 + m - { n\over 2} + s )
\over s! } ( 1- { 1 \over x} )^{ s + 1 - n }\ ,
}
after using \identu\ with $\tilde m =1$ and $\tilde n = n-1$.
The result for $\chi$ is therefore
\eqn\sumev{\eqalign{
\hat \chi = & { \Gamma(n-1)} \sum_{s=0}^{n-2} { 1 \over s! }
( 1- { 1 \over x} )^{ s + 1 - n } Y(s,n,y)\ ,
\cr
Y(s,n,y) \equiv & \sum_{m=0}^\infty {   \Gamma(2 + m - { n \over 2 } + s)
\Gamma(m + { n \over 2 }) \over
\Gamma(2 + m - { n \over 2} )^2 } y^{-m }
\cr
=&{ \Gamma(2 - { n\over 2 } + s) \Gamma( { n\over 2})
\over \Gamma( 2 - { n\over 2} )^2}   ~_3F_2( 1,2 - { n\over 2 } + s,{ n\over 2};
 2 - { n\over 2} , 2 - { n \over 2} ; 1/y )\ .
 }}
We now use the identity
\eqn\fttwoid{
 ~_3F_2(a_1,a_2,a_3; b_1,b_2,z)  ={ \Gamma(b_2) \over \Gamma(a_3) \Gamma(b_2 - a_3) } \int_0^1
 dt t^{a_3 -1} (1-t)^{b_2 -a_3-1} ~_2F_1(a_1,a_2,b_1; t z )\ .
 }
Applying this formula and choosing $a_1=1$, we find that the parameters of the hypergeometric
function ${}_2F_1$ are related in such a way that we can apply Eqn.~\identu\ with
\eqn\newis{
\tilde m =  a_1=1\ ,
\qquad
\tilde n =  a_1+ a_2 - b_1 = 1 + s\ ,
\qquad
a_3 =  {n\over 2} \ .
}
(we take $n$ to be a real number and $s$ to be integer). We then find
\eqn\redfsi{\eqalign{
{}_3F_2( 1,2) & - { n\over 2 } + s,{ n\over 2};
 2 - { n\over 2} , 2 - { n \over 2} , 1/y ) =
 \cr
~~~~ =& { \Gamma( 2 - {n\over 2} )^2 \over \Gamma({n \over 2 }) \Gamma( 2- n) }
 \int_0^1
 dt \, t^{{n\over 2} -1} (1-t)^{1-n}   { 1 \over \Gamma(2 - { n \over 2 })
 } {}_2F_1(1,2-{n\over 2 } + s, 2 - { n\over 2}, t/y)
 \cr
~~~~ =& { \Gamma( 2 - {n\over 2} )^2 \over \Gamma({n \over 2 }) \Gamma( 2- n) }
 \int_0^1
 dt \, t^{{n\over 2} -1} (1-t)^{1-n}   { \Gamma(s+1) \over \Gamma(1-{n\over 2}) \Gamma(2 - { n \over 2} +s)}
\cr
& \qquad \qquad
\times \sum_{q=0}^s {\Gamma(1 - { n\over 2} + q) \over q! } (1-t/y)^{q - s -1}
\cr
~~~~ = &{ \Gamma( 2 - {n\over 2} )^2 \over \Gamma({n \over 2 }) \Gamma( 2- n) }
   { \Gamma(s+1) \over \Gamma(1-{n\over 2}) \Gamma(2 - { n \over 2} +s)}
\cr
& \qquad \qquad \times \sum_{q=0}^s {\Gamma(1 - { n\over 2} + q) \over q! } \int_0^1
 dt \, t^{{n\over 2} -1} (1-t)^{1-n} (1-t/y)^{q - s -1}\ ,}
}
and
\eqn\isint{
 \int_0^1
 dt \, t^{{n\over 2} -1} (1-t)^{1-n} (1-t/y)^{q - s -1} =
 {  \Gamma({n \over 2} ) \Gamma(2-n) \over
    \Gamma(2 -{ n \over 2})} ~_2F_1({n\over 2} , 1-q + s, 2 - { n\over 2} ; 1/y)\ .
 }
Let us now go back to the sum we wanted to evaluate \sumev:
\eqn\simpf{
Y(s,n,y) = {   \Gamma( { n\over 2})
\over   \Gamma( 2- {n \over 2}) }
   { \Gamma(s+1) \over \Gamma(1-{n\over 2}) } \sum_{q=0}^s {\Gamma(1 - { n\over 2} + q) \over q! }
~_2F_1({n\over 2} , 1-q + s, 2 - { n\over 2} ; 1/y)\ .
}
Using Eqn.~\identu, with  $\tilde m = 1- q + s$, $b=n/2$ and $\tilde n = n-1 + s - q$,  we obtain
\eqn\simpfte{\eqalign{
Y(s,n,y) = &
   { \Gamma(s+1) \Gamma(n-1) \over \Gamma(1-{n\over 2})}
   \sum_{q=0}^s {\Gamma(1 - { n\over 2} + q)  \over q!  \Gamma(1 + s -q) \Gamma(-n/2 + 1 -s +q) }
 \cr
  & \times \sum_{k=0}^{n-2} { \Gamma( - n/2 + 1 - s + q + k) \Gamma(n-1 + s -q  - k)
 \over k! \Gamma(n-1 -k)} \left( 1- { 1 \over y}\right)^{ k -n + 1 + q -s }\ .
}}
So the final answer takes the form
\eqn\chihfin{\eqalign{
\hat \chi (x,y) =&
{\Gamma(n-1)^2 \over \Gamma( 1- { n\over 2} ) } \sum_{s=0}^{n-2} \sum_{q=0}^s
\sum_{k=0}^{n-2}
 {\Gamma(1 - { n\over 2} + q)  \over q!  \Gamma(1 + s -q) \Gamma(-n/2 + 1 -s +q) }
 \cr
  &   \times { \Gamma( - n/2 + 1 - s + q + k) \Gamma(n-1 + s -q  - k)
 \over k! \Gamma(n-1 -k)}
\cr
& \times \left( 1- { 1 \over y}\right)^{ k -n + 1 + q -s } \left(1-{1 \over x}\right)^{s + 1 -n }\ ,
}}
with
\eqn\chiiss{\eqalign{
\chi = &
- {\zeta(n) \over (-2 \pi)^n }{ 1 \over x_1 x_2^2} { \Gamma(n-1)  \over \Gamma(1-{n \over 2} )}
\left[ g \left( 1 - { 1 \over x_2^2}\right)\left(1 - { 1 \over x_1x_2}\right) \right]^{1-n}
\cr
& \qquad \times \sum_{s=0}^{n-2} \sum_{q=0}^s \sum_{k=0}^{n-2}
 {\Gamma(1 - { n\over 2} + q)  \over q!  \Gamma(1 + s -q) \Gamma(-{ n \over 2} + 1 -s +q) }
 \cr
&  \qquad \times { \Gamma( - { n \over 2} + 1 - s + q + k) \Gamma(n-1 + s -q  - k)
 \over k! \Gamma(n-1 -k)} \left({ 1 - { 1 \over x_1x_2} \over{  1- { 1 \over x_2^2}}} \right)^s
  \left(1-{1 \over x_2^2}\right)^{k +q }\ .
}}

Note that we have expressed the full phase in the second expression.
 The leading-order answer in the
scaling limit comes from $q=k=0$. In that case one can see that this expression reduces to
Eqn.~\fullnoth , after a relabeling of the indices.
When $n$ is even, we can rewrite the last result in a way that is slightly more concise:
\eqn\chihfin{\eqalign{
\chi = &- {\zeta(n) \over (-2 \pi)^n }{ 1 \over x_1 x_2^2} { \Gamma(n-1) \Gamma({n\over 2} ) }
\left[ g ( 1 - { 1 \over x_2^2})(1 - { 1 \over x_1x_2}) \right]^{1-n}  \times \cr
& \sum_{s=0}^{n-2} \sum_{q=0}^s
\sum_{k=0}^{n-2} (-1)^{q+k}
 {\Gamma( { n\over 2} + s - q)  \over q!  \Gamma(1 + s -q) \Gamma({ n \over 2} -q) }\times
 \cr
  &   {  \Gamma(n-1 + s -q  - k)
 \over \Gamma(  { n \over 2}  + s - q - k) k! \Gamma(n-1 -k)} \left({ 1 - { 1 \over x_1x_2} \over{  1- { 1 \over x_2^2}}} \right)^s
  (1-{1 \over x_2^2})^{k +q }\ .
}}

%%%%%%%%%%%%%%%%%%%%%%%%%%%%%%%%%%%%%%%%%%%%%%%%%%%%%%%%%%%%%%%%%%%%%%%%%%%%
\appendix{D}{Comments on the analytic structure of the phase}
\noindent
%%%%%%%%%%%%%%%%%%%%%%%%%%%%%%%%%%%%%%%%%%%%%%%%%%%%%%%%%%%%%%%%%%%%%%%%%%%%
We consider the antisymmetric combination
\eqn\combin{
 \chi_a(w_1,w_2) =  \chi(w_1,w_2) -   \chi(w_2,w_1)\ ,
}
since $\chi$ always appears in this combination.
We then define $w_c \equiv w_1 + w_2$ and $w_r \equiv w_1 - w_2 $, and add the tree-level
result in Eqn.~\afsans\ to the above expression.  We define by
$\chi'_a$ the antisymmetric combination of the form in Eqn.~\combin, including
all contributions except that at one-loop order.
Up to terms that will cancel in the full phase, we find
\eqn\anscomb{\eqalign{
\partial_{w_r} \chi'_{ a} =&
 - \int_0^\infty   d \tau \left\{{ \tau  \over \tau + w_c }
\left[ { 1 \over  e^{ 2 \pi \tau (\tau + w_c - w_r) } -1}  +
{ 1 \over  e^{ 2 \pi \tau (\tau + w_c + w_r) } -1} +1
  \right] \right.
  \cr
  & \left. -1 + { w_c \over \alpha + \tau } \right\}\ ,
  }}
where the last two terms are introduced to make the integral finite. The quantity $\alpha$ is a constant
that will drop out of the combinations appearing in the total phase \contras .  In the end we find that
these last two terms can essentially be ignored.

We are interested in understanding the analytic structure of this function. We see that the integrand
is an analytic function of $\tau$ with poles at values that depend on $w_c,~w_r$.
Let us first understand the branch points.  These will arise when any of the singularities
of the integrand approaches $\tau =0$. In general, if we have an integral of the form
\eqn\integr{
\int_0^\infty d\tau { 1 \over \tau - a} h(\tau)\ ,
}
then we find a branch cut at $a=0$ of the form
\eqn\branchu{
h(a)  \log a \ .
}
Following this logic, one might expect that there is a branch cut at $w_c=0$. This branch cut
is actually not present because the residue on this pole vanishes.  This happens because we combined
the $n\geq 2$ contribution with the tree-level answer \afsans .
This shows that the branch cut at
$w_1 + w_2$, which is present in the tree-level answer, is removed in the exact answer.
Note also that
the one-loop answer does not have a branch cut at this position. This, in turn, implies that the
branch cut in \scattp\ is  removed.

We can then look at points where the exponentials in \anscomb\ lead to poles in the integrand.
These occur at $\tau = \tau^\pm_n(w_1), ~\tau^\pm_n(w_2)$, where
\eqn\brptau{
\tau^\pm_n(w_2) = -w_2 \pm \sqrt{ w_2^2 + in }\ .
}
In these cases the residues do not vanish, and we find branch cuts that
arise when we move along a contour in $w_2$, in such a way that
$\tau^\pm$ encloses the origin.  If $\tau^+$ circles around the origin in an anticlockwise
manner, then we get the following shift in the phase:
\eqn\brataup{
 \Delta \chi'_{a}|_{\tau^\pm_n \to e^{2 \pi i } \tau_{n}^\pm }
  \sim i  \log( w_1 + w_2 + \tau_n^\pm(w_2) )\ .
 }
  We find similar terms with $w_2 \to w_1$, and a flip of the overall
 sign.  Note that we have integrated the result obtained from analyzing the singularities in
  Eqn.~\anscomb .

The case corresponding to $\tau^-_0(w_2) = - 2 w_2$ deserves special attention, since its partner
$\tau^+_0 =0$ apparently does not encircle the origin. In this case we get
\eqn\gettauz{
\Delta \chi(w_1 , w_2^+)|_{w_2^+ \to e^{2 \pi i } w_2^+ } = - i \log \left( { w_1 + w_2^+
\over w_1 - w_2^+ } \right)\ .
}
We get a similar expression for $w_2^-$ enclosing the origin:
\eqn\gettauz{
\Delta \chi(w_1 , w_2^-)|_{w_2^- \to e^{2 \pi i } w_2^- } = - i \log \left( { w_1 + w_2^-
\over w_1 - w_2^- } \right)\ .
}
If one thought that this was the only contribution obtained from double crossing, then
one would find a contribution coming from the one-loop answer that precisely cancels this.

However, if one follows the path in  \contours , then there is an additional contribution
arising in $\chi(w_1,w_2^+)$ from the term involving $\tau^\pm_{-1}(w_2^+) =
- w_2^+ \pm w_2^- $.
We find that $\tau^+_{-1}(w_2^+) = -w^+_2 +w_2^-$ encircles the origin in a clockwise fashion,
while $\tau^-_{-1}(w_2^+)=-w_2^+- w_2^-$ cycles around the origin in an anticlockwise fashion.
This can be seen most easily by using the variable $\eta$ in \wpmfrometa, so that
$ - w_2^+ \pm w_2^- \sim e^{ \mp \eta }$, while remembering that $\eta \to \eta + 2 \pi i $
under the double-crossing transformation.
The net contribution from these two terms is
\eqn\chishif{
\Delta \chi(w_1, w_2^+) |_{\tau^\pm_{-1} \to e^{ \mp 2 \pi i } \tau^\pm_{-1}}=
 - i \log\left( { w_1 + w_2^-
\over w_1 - w_2^- } \right)\ .
}

When we consider the contour for $w_1^-$, we find that $\tau_{1}^\pm(w_1^-) = - w_2^- \pm w_2^+$
 enclose the origin. In fact, these are the same combinations that we considered above,
producing a contribution of the form
\eqn\chishif{\eqalign{
& \Delta \chi(w_1, w_2^-) |_{\tau^\pm_{1} \to e^{ \mp 2 \pi i } \tau^\pm_{1}}=
 - i \log\left( { w_1 + w_2^+
\over w_1 - w_2^+ } \right)\ .
 }}

With the contour choice we have made, all the other $\tau^\pm_{n}$, which were
not explicitly considered, do not encircle the origin and therefore do not give rise to a
shift in $\chi$.
We see that when we add all of these contributions together, the shift of
$\Delta [ \chi(w_1,w_2^+) - \chi(w_1,w_2^-)]$ vanishes. Here we will have to sum
over $w_1 \to w_1^\pm$, but each one vanishes on its own.
Thus we see that, with the contour
choice that we have made, the double crossing equation is obeyed.  Note, however, that
the branch cuts at $w^\pm=0$ in the one-loop expression seem to be canceled by
the branch cut coming form higher-order contributions.  Of course,
it would be nice to understand the analytic properties of the full phase
more completely.

%%%%%%%%%%%%%%%%%%%%%%%%%%%%%%%%%%%%%%%%%%%%%%%%%%%%%%%%%
\subsec{A check of the single crossing equation.  }
\noindent
%%%%%%%%%%%%%%%%%%%%%%%%%%%%%%%%%%%%%%%%%%%%%%%%%%%%%%%%%
 The contribution to the phase from the terms with $n\geq 2 $ can be written as
 \eqn\fullfa{\eqalign{
 \theta = & { 1 \over 2 \pi }  \int_{w_2^+}^{w_2^-}
 { d \tau} ( { 1  \over \tau + w_1^+}  - {1 \over \tau + w_1^-})
 \log\left[ { 1 - e^{ - 2 \pi (\tau^2 -  (w_2^+)^2  ) }
  }\right]-
 \cr
 &{ 1 \over 2 \pi }  \int_{w_1^+}^{w_1^-}
 { d \tau} ( { 1  \over \tau + w_2^+}  -{1 \over \tau + w_2^-})
 \log\left[ { 1 - e^{ - 2 \pi (\tau^2 -  (w_1^+)^2  ) }
 }\right]\ ,
 }}
where we have used that $e^{ 2 \pi (w^+)^2} = e^{2 \pi (w^-)^2 }$.
To check the crossing relation, we want to evaluate $\bar \theta + \theta \equiv
\theta(-w_1, w_2) + \theta(w_1,w_2)$.
Let us first ignore possible terms that are acquired under continuation.
We therefore write this expression as follows:
\eqn\evaluc{\eqalign{
\bar \theta + \theta = &
{ 1 \over 2 \pi }  \int_{w_2^+}^{w_2^-}
 { 2 \tau d \tau} ( {1  \over \tau^2 - ( w_1^+)^2}  - {1 \over \tau^2 - (w_1^-)^2})
 \log\left[ { 1 - e^{ - 2 \pi (\tau^2 -  w_2^2  ) }
 }\right]
 \cr
 & - { 1 \over 2 \pi }  \int_{w_1^+}^{w_1^-}
 { 2 \tau d \tau} ( { 1  \over \tau^2 - (w_2^+)^2}  -{1 \over \tau^2 - (w_2^-)^2})
 \log\left[ { 1 - e^{ - 2 \pi (\tau^2 -  w_1^2  ) }
 }\right]\ ,
 }}
where we have combined the term scaling like $1/(\tau + w_1^+)$ from $\theta$ with
the term scaling like $1/(\tau - w_1^+)$ from $\bar \theta $.
In the second line we see that
$\bar \theta$ involves an integral in the interval $[-w_1^+,-w_1^-]$.
Under a change of integration variables $\tau \to - \tau$,
we recover an integral with the same limits as the one we had in $\theta$,
but with an extra minus sign in the denominator.
It is assumed that in all of these manipulations we do not cross the contour
or acquire any extra contributions to the expressions.

We can now invoke a change of variables $u = \tau^2 - (w_2^+)^2$ in the first integral,
along with a similar change in the second integral.
Imposing the constraint between $w_i^\pm$ in Eqn.~\constrbe, we get
\eqn\evaluctr{\eqalign{
\bar \theta + \theta = &
{ 1 \over 2 \pi }  \int_{0}^{-i}
 { du } ( {1  \over u - \delta}  - {1 \over u -\delta + i })
 \log\left[ { 1 - e^{ - 2 \pi u }
 } \right]-
 \cr
 &{ 1 \over 2 \pi }  \int_{0}^{-i}
 { d u } ( { 1  \over u + \delta }  -{1 \over u + \delta + i })
 \log\left[ { 1 - e^{ - 2 \pi u }
 }\right]\ ,
\cr
\delta \equiv & (w_1^+)^2 - (w_2^-)^2\ .
 }}
Note that $Re(\delta ) \ll  0$ in the region $|w_2^+| \gg |w_1^\pm |$, with $w_2^+$ almost real.
By changing variables $u' = - ( u+i)$ in the second line,
we find that the limits of integration
of the second term are the same as in the first term.
In addition, the integrand looks very similar.
After combining terms we obtain
\eqn\eluct{\eqalign{
\bar \theta + \theta = &
{ 1 \over 2 \pi }  \int_{0}^{-i}
 { du } ( {1  \over u - \delta}  - {1 \over u -\delta + i })
( \log\left[ { 1 - e^{ - 2 \pi u }
} \right]-\log\left[ { 1 - e^{  2 \pi u }
 }\right])\ ,
}}
where
\eqn\logis{
\log\left[ { 1 - e^{ - 2 \pi u } } \right]-\log
 \left[ { 1 - e^{  2 \pi u } }\right] = -2 \pi u - i \pi\ .
 }
 Note that the $- i \pi$ is correct in the difference of logs, since the first log is defined
such that it is real for $u>0$, and the second so that it is real for $u<0$.
We then get
 \eqn\finsec{
\bar \theta + \theta  =  -(\delta + i/2) \log { -i - \delta \over - \delta } + ( \delta - i/2)
 \log { -\delta \over - \delta + i }\ .
}

We now compute the change in the tree level-expression \afsans:
\eqn\afcross{
\bar \theta^0 + \theta^0 =
- 2 \delta \log ( - \delta) - ( -\delta +i ) \log (-\delta + i) - (-\delta - i ) \log (- \delta -i )\ .
}
When we add this to the results of Eqn.~\eluct\ we find
\eqn\chite{
( \bar \theta + \bar \theta^0) + ( \theta + \theta^0)
= - { i \over 2 } \log { -\delta + i \over - \delta - i }\ .
}
After taking into account the change in the one-loop answer, the total change
should indeed be \chite\ (see \bhl\ for further details).

%%%%%%%%%%%%%%%%%%%%%%%%%%%%%%%%%%%%%%%%%%%%%%%%%%%%%%%
\subsec{Poles}
\noindent
%%%%%%%%%%%%%%%%%%%%%%%%%%%%%%%%%%%%%%%%%%%%%%%%%%%%%%%
Let us now focus on the poles in the amplitude.  It is simplest to consider the expression
involving a derivative \anscomb , and the poles in the $S$ matrix will give rise to
poles in Eqn.~\anscomb .  These poles
 arise when two singularities of the integrand come close together. In principle, they
 have to pinch the integration contour, but it seems that by analytically continuing
in a suitable manner we will get a singularity on some of
 the branches discussed above.

 An interesting set of singularities arises when
 \eqn\signar{
  -w_c = \tau_n^\pm(w_2) , ~~~~~{\rm or}~~~ 0= w_1 \pm \sqrt{(w_2)^2 + i n }\ .
  }
 At these singularities we find that $w_r w_c = i n$.  We can therefore take $w_r = in/w_c + \epsilon$ and
 expand the integral for small $\epsilon$.  When we do this we see that the contour can be pinched, and
 we can find a singularity in the phase of the form
 \eqn\singph{
 \chi'_{a } = - i \log \epsilon \sim - i \log( w_1 \pm \sqrt{ (w_2)^2 + i n} )\ .
 }

 The analysis thus far does not immediately tell us whether we get a singularity in the
 region close to the physical space.  We expect that some general
 principle will tell us what the physical region
 is.  For the moment, we will just define it as having values of $w^\pm$,
 which are close to the physical values and far from the branch points.

 In particular, we can consider the scattering amplitude in the giant-magnon region, where
 $k_i \gg 1$.  In this regime, $ w_i $ are large and almost purely imaginary,
  and we find that all branch points are far from the
  physical line.  We would then like to know if there are any poles that arise as we analytically
  continue around this region by a small amount.  For this to happen, we need to have values
  of $\tau^\pm_n(w_2)$ and $w_1 + w_2$  that can pinch the contour.  In particular, we would like to
  have $w_1 + w_2 $ be small.  This happens only if we consider the combinations
  $(w_1^+,w_2^-)$ and $(w_1^-,w_2^+)$.  We can then find that when we analytically continue
 in the region where $k_1 \sim k_2 \gg 1$, we pinch the contour and get poles that,
  by our definitions, are close to the physical line.

  A simple way to see this is to start with the expression in Eqn.~\anscomb, for the case with
  $w_1 = w_1^-$ and $w_2 = w_2^+$, with $k_1,~k_2 \gg 1$.
   We can then rotate the contour to the line
  $\tau = e^{ i {\pi \over 4}} y$, with real and positive $y$. The integral along the
  new contour gives an  exponentially small contribution
   (up to analytic terms in the phase).
  So the only remaining
 contribution arises from the poles that we pick up when we rotate the contour.
 These poles are at
 \eqn\taupolpi{
 \tau = \tau^+_n(w_2^+) = -w_2^+ + \sqrt{(w_2^+)^2 + i n }~,~~~~~~n>0\ ,
} where the branch of the square root is chosen so that, for
physical values of $w^+$, the real part of the square root is
positive. The contribution at each pole is given in terms of the
residue of the integrand, and is equal to \eqn\polecontr{ -i {
\tau^+_n(w_2^+) \over \tau^+_n(w_2^+)+ w_2^+ + w_1^-} { 1 \over 2 (
\tau^+_n(w_2^+) + w_2^+)} = - \partial_{w_r} i \log( \tau^+_n(w_2^+)
+ w_2^+ + w_1^- )\ . } There is a similar contribution from the
terms with $(w_1^+,w_2^-)$. These two combine to give
\eqn\combinps{\eqalign{
 \sigma^2 \sim e^{ - 2 i
\chi'_a(w_1^-,w_2^+) + 2 i \chi'_a(w_2^-,w_1^+) } & \sim
\prod_{n=1}^\infty {  (
\sqrt{(w_1^+)^2 + i n } + w_2^- )^2 \over ( \sqrt{(w_2^+)^2 + i n } + w_1^- )^2 }\ ,
\cr
 \sim & \prod_{n=1}^\infty {  (
\sqrt{(w_1^+)^2 + i n } + w_2^- ) ( \sqrt{(w_2^-)^2 - i n } + w_1^+ )\over
( \sqrt{(w_2^+)^2 + i n } + w_1^- ) ( \sqrt{(w_1^-)^2 - i n } + w_2^+ )\ .
}
}}
 The products in the right hand side are defined only up to
   analytic terms that are divergent.\foot{ To produce
a formula that makes mathematical sense, we can take the log on
each side and take derivatives $\partial_{w_1^+}
\partial_{w_2^+}$.  The divergence comes from a similar divergent
term that we dropped after we rotated the contour.  The expression
we started from, on the left-hand side, is finite.}
 We have written a couple of different
expressions whose poles are the same in the region of interest.   This
expression is similar to the one written for the so called ``giant
magnon'' phase in \bhl . In fact, one can add other contributions
to obtain an approximate expression that looks like the ``giant'' guess in
\bhl . However, the full phase is
not the same as the ``giant magnon'' phase in \bhl , since the expression
\fullch\ has a branch cut at $w_2^\pm =0$, while the expression in
the ``giant magnon'' phase in \bhl\ does not. In terms of the
variable $u$ introduced in \varu, the poles lie at \eqn\plesu{ u_1
- u_2 = i n ~,~~~~~~~~~n>1\ . } Notice that displaying the poles
as in \plesu\ is misleading, since in the region $u_i \gg 0$ we do
not have poles near the physical region. This is the plane-wave
region. On the other hand, for $u_i \ll 0$, which is the
giant-magnon region, these poles are close to the physical region.
The reason that this happens is that there are many branch cuts
starting from the one at $u_i = \pm i/2 $ (see \poles ).

It turns out that at $u_1 - u_2 = i$, the factor $\sigma^2$ is finite, with no zeros or poles.
We should simply  note that \finantisym\ is analytic at $w_1+ w_2 =0$.
The following is a longer argument demonstrating this fact.
To analyze possible poles at $u_1 - u_2 =i$, or, more precisely,
at $w_1^+ + w_2^- =0$,
 we need to analyze the one loop-term in Eqn.~\simplestsol .
So far we have not discussed the one-loop term because, due to its form in \hgunf, we see
that it can produce only poles or zeros for $u_1-u_2 = 0$ or $u_1 - u_2 =\pm i$.
Most of these poles are in
 other branches that one reaches after going through the
cuts at $u_i = \pm i/2$.  We are only going to discuss poles in the main branch, which is
directly connected to the physical values of $u$.
For this purpose, it is useful to notice that as we change the physical momentum $k$
from zero to infinity, the imaginary part of $\hat \theta^+ $ in Eqn.~\newlim\ goes between zero
and $- i \pi/2$.  Similarly, the imaginary part of $\hat \theta^- $ goes between zero and $+ i \pi/2$.
When we do an analytic continuation in the neighborhood of the physical values of $w^\pm$, the
imaginary part of $\hat \theta^\pm$ will not change by a large amount.
Since differences between $\hat \theta^\pm_i$ can only be at most $\pm i \pi$, we only need to
worry about the following terms in $h$ (see \hgunf ):
\eqn\twotermw{
h( \theta) \sim { ( \theta + i \pi ) \over ( \theta - i \pi ) }\times \cdots
}
In addition, the only differences that can get close to $\pm i \pi$ are
$\theta_1^+ - \theta_2^- \sim - i \pi $ or $\theta_1^- - \theta_2^+ \sim i \pi $.
In the first case we see from Eqns.~\twotermw\ and \simplestsol\ that we will get a pole in
$\sigma^2$ as
$w_1^+ + w_2^-\sim 0 $.  In the second case we get a zero in $\sigma^2 $ as
 $w_1^- + w_2^+ \sim 0$.

When we analyzed the contributions from the higher loop terms coming from \anscomb\ we
considered derivatives with respect to $w_r = w_1 - w_2$ for fixed $w_1+w_2$, so we could
have missed contributions involving just $w_1+w_2$. To find such contributions
from the higher loop terms, we would be tempted to simply set $w_1=w_2$ first (which amounts
to setting $w_r =0$) and then we would find that there is no singularity at $w_c =0$.
However, we should be more careful, since the function has branch cuts at $w_i=0$.
It is safer to keep $w_r$ finite, as it would be in the physical (giant magnon) 
region we are exploring,
and then to compute
the behavior of the function for small $w_c$. Approximating the integral for small $w_c$,
we find that singularities arise from small $\tau$.  We may thus expand the integral
for small $\tau$ to find
\eqn\newchip{
\chi_a' \sim  { 1 \over 2 \pi } \int { d \tau \over \tau + w_c } \log
\left[ { \tau - w_r \over \tau + w_r} \right] = - { i \over 2 } \log w_c \ ,
}
where we had to use that
\eqn\usefim{
\log { - w_r \over w_r } = \log { w_2^+ - w_1^- \over w_1^- - w_2^+ } \sim i \pi \ ,
}
when we start from physical particles and then move to the limit where $w_2^+ + w_1^- \sim 0$.
Thus we see that the higher loop contributions to $\sigma^2 $ produce a single pole of the form
$\sigma^2 \sim (w_2^+ + w_1^-)^{-1}$. This pole cancels the one-loop term and gives no
net pole or zero in $\sigma^2$ for $w_2^+ + w_1^- \sim 0$.

\listrefs
\bye